\documentclass{aa}
\usepackage{graphicx}
\usepackage{txfonts}
\usepackage{lscape}
\begin{document}
   \title{Multi-wavelength study of X-ray luminous clusters at $z \sim 0.3$
}

   \subtitle{I. Star formation activity of cluster galaxies}

   \author{F. G. Braglia\inst{1,3}
          \and
          D. Pierini\inst{1}
	  \and
	  A. Biviano\inst{2}
	  \and
	  H. B\"{o}hringer\inst{1}
          }

   \offprints{F. G. Braglia}

   \institute{Max-Planck-Institut f\"{u}r extraterrestrische Physik, 
              Giessenbachstrasse 1, D-85748 Garching bei M{\"u}nchen, Germany\\
              \email{fbraglia@mpe.mpg.de; dpierini@mpe.mpg.de; hxb@mpe.mpg.de}
	      \and
              INAF-Osservatorio Astronomico di Trieste,
              via G. B. Tiepolo 11, I-34143 Trieste, Italy\\
	      \email{biviano@oats.inaf.it}
	      \and
              University of British Columbia, Dept. of Physics and Astronomy,
              6224 Agricultural Road, V6T 1Z1 Vancouver BC, Canada\\
	      \email{fbraglia@phas.ubc.ca}
             }

   \date{}

 
  \abstract
   {The current paradigm of cosmic formation and evolution of galaxy clusters foresees growth mostly through merging. Galaxies in the infall region or in the core of a cluster undergo transformations owing to different environmental stresses.}
   {For two X-ray luminous clusters at redshift $z \sim 0.3$ with opposite X-ray morphologies (i.e., dynamical states), RXCJ\,0014.3-3022 and RXCJ\,2308.3-0211, we assess differences in galaxy populations as a function of cluster topography. This is a pilot study for the joint X ray and optical analysis of the REFLEX-DXL cluster sample.}
   {Cluster large-scale structure and substructure are determined from the combined photometry in the B, V, and R bands, and from multi-object optical spectroscopy at low resolution. Photometric redshifts and broad-band optical colours are determined. A spectral index analysis is performed, based on the [O\,II]($\lambda \lambda 3726, 3728$\,\AA) and $\mathrm{H_{\delta}}$($\lambda 4102$\ \AA) features, and the $D4000$ break, which are available for more than 100 member galaxies per cluster. Additional far-ultraviolet (FUV) photometry is retrieved from the GALEX archive. Combination of spectral indices and FUV--optical colours provides a picture of the star formation history in galaxies.}
   {In spite of the potential presence of a small fraction of galaxies with obscured star formation activity, the average star-formation history of cluster members is found to depend on cluster-centric distance and, more interestingly, on substructure. The core regions of both clusters mainly host galaxies dominated by old, passively evolving stellar populations, which define the same red sequence in a $(B-R)$ colour--$R$ magnitude diagram. However, there is a sharp increase in star formation activity along two well-defined filamentary structures of the merging cluster RXCJ\,0014.3-3022, out to its virial radius and beyond. It is produced by luminous (i.e., with $L_\mathrm{R} \geq L_\mathrm{R}^{\star}$) and sub-L$^{\star}$ galaxies. Conversely, the regular cool-core cluster RXCJ\,2308.3-0211 mostly hosts galaxies which either populate the red sequence or are becoming passive. This holds out to its virial radius, and also for its large-scale environment.}
   {These results suggest the existence of a correspondence between assembly state and overall age of the stellar populations of galaxies inside the virialized region and in the surrounding large scale structure of massive clusters at $z \sim 0.3$.}


   \titlerunning{Substructure and star formation activity of galaxies in massive clusters}
   \maketitle
%

\section{Introduction}\label{intro}

X-ray emitting clusters signpost the largest amounts and concentrations
of (cold) dark matter (CDM) in bound systems.
At the same time, they provide large samples of coeval galaxies
within a well-defined environment.
This enables to study how galaxy properties, like star formation rate (SFR),
behave as a function of environment and local density.

Detailed numerical simulations confirm that clusters tend to form
at the intersection of filaments and sheets
in the evolving large-scale structure of the Universe:
over time, matter falls along such structures and accretes into a cluster.
The infall pattern is not random: paths show a correlation in time
and are almost stable, so that matter is channelled into a cluster
through filamentary structures with persistent geometry.
Matter does not accrete in a steady and continuous way,
but in clumps and bound structures, which can be identified
as infalling groups of galaxies or less massive clusters
(Colberg et al.~\cite{colberg99}).

The existence of extended filamentary structures and voids
in the large-scale spatial distribution of local galaxies
is known since the advent of redshift surveys (e.g. Gregory and Thompson~\cite{gregory78};
Davis et al.~\cite{davis82}).
Deep X-ray imaging of the emission from hot ($10^5$--$10^7~\mathrm{K}$) gas
and analysis of the two-dimensional distribution of galaxies
at the same (photometric or spectroscopic) redshift of a cluster
have extended the detection of filamentary networks around clusters
up to redshift $z \sim 0.6$ (e.g. Kull \& B\"{o}hringer~\cite{kull99};
Scharf et al.~\cite{scharf00}; Kodama et al.~\cite{kodama01};
Zappacosta et al.~\cite{zappacosta02}; Durret et al.~\cite{durret03};
Ebeling et al.~\cite{ebeling04}; Dietrich et al.~\cite{dietrich05}).

The galaxy populations in cluster-feeding filaments can be expected to differ
somewhat from those in the cluster main body or in the adjacent field,
as filaments are an environment of transition between the low
and extremely high density regimes.
Consistently, observations have revealed a clear correlation
between galaxy morphology or SFR and local density or cluster-centric distance
(Dressler et al.~\cite{dressler80}; Dressler et al~\cite{dressler97}; Whitmore et al.~\cite{whitmore93};
Balogh et al.~\cite{balogh97}).
The past and present star-formation activities of galaxies
appear to be different in low and high density environments
already at intermediate redshifts (Abraham et al.~\cite{abraham96};
Morris et al.~\cite{morris98}; Poggianti et al.~\cite{poggianti06}).
From a spectral index analysis of galaxies in 15 X-ray luminous clusters
at $0.18 < z < 0.55$, Balogh et al. (\cite{balogh99}) concluded that
the increase in the observed star-formation activity towards the outer regions
of clusters is consistent with an age sequence: galaxies in cluster outskirts
have experienced star-formation phases more recently than galaxies
in cluster cores.
Their interpretation was that the truncation of star formation
in cluster galaxies may largely be a gradual process,
perhaps due to the exhaustion of gas in galaxy discs
over fairly long timescales.
In this case, differential evolution may result
because field galaxies can refuel their discs with gas from extended halos,
thus regenerating star formation, while cluster galaxies may have lost
such gaseous halos as well as the dark matter ones (cf. Halkola et al.~\cite{halkola07}),
and so continue to evolve passively.

Combining X-ray and optical observations, B\"{o}hringer at al.
(\cite{boehringer06}) and Braglia et al. (\cite{braglia07},
hereafter referred to as BPB07) provided evidence for the presence
of two filaments out from the main body
of the cluster RXCJ\,0014.3$-$3022 (alias Abell~2744, AC\,118)
and extending beyond its virial radius.
This is one of the 13 clusters at $z=0.27$--0.31 with X-ray luminosity
higher than $10^{45}~\mathrm{erg~s^{-1}~cm^{-2}}$, selected from
the ROSAT-ESO Flux Limited X-ray (REFLEX) cluster survey
(B\"{o}hringer et al.~\cite{boehringer01a})
and observed with XMM-\emph{Newton}, which make the Distant X-ray Luminous
(DXL) cluster sample (Zhang et al.~\cite{zhang06}).
It is also a major-merger system (e.g. Boschin et al.~\cite{boschin06})
and a ``Butcher--Oemler'' cluster (Butcher \& Oemler~\cite{butcher78a},
\cite{butcher78b},~\cite{butcher84}).
The blue galaxies in the cluster core, that are more abundant
compared to present-day massive clusters, are mostly systems
involved in major mergers but sub-L$^{\star}$ even in this brightened phase;
in their faded state they appear destined to become dwarfs
(Couch et al.~\cite{couch98}).
In addition, BPB07 found evidence for luminous galaxies
along the two filaments which exhibit a boosted star-formation activity
with respect to their counterparts in the field.

More recent studies seem to confirm the increase of star formation activity
along filaments in the outskirts of clusters.
For instance, Fadda et al. (\cite{fadda08}) report the presence
of twice as many dusty starbursts along two filamentary structures
connecting the clusters Abell~1770 and Abell~1763 ($z=0.23$)
as in other cluster regions.
Furthermore, Porter et al. (\cite{porter08}) find a significant enhancement
of star formation, mostly in dwarf galaxies ($M_\mathrm{B}>-20$),
from a sample of 52 supercluster-scale filaments of galaxies
joining pairs of rich clusters of galaxies
within the 2-degree Field Redshift Survey region.
This work consolidates a similar result
obtained for the Pisces-Cetus Supercluster filaments
(Porter \& Raychaudhury~\cite{porter07}).

The presence of dwarf and giant galaxies with enhanced SFR
and the wealth of available data make the merging cluster RXCJ\,0014.3-3022
an interesting laboratory to study how different physical processes
affect galaxy evolution in different cluster environments.
To better understand the importance of substructure on star-formation activity
in cluster galaxies, we investigate the DXL cluster RXCJ\,2308.3-0211
(alias Abell~2537), a cool core (thus relaxed) system,
in addition to RXCJ\,0014.3-3022.
The analysis presented here shows the power of combining X-ray observations,
wide-field optical imaging, and multi-object spectroscopy for such a study
(see also B\"{o}hringer et al.~\cite{boehringer01b},~\cite{boehringer06},
\cite{boehringer07}; BPB07; Pierini et al.~\cite{pierini08}).
Results from a similar analysis applied to the entire REFLEX-DXL sample
will be presented in future papers of this series.

This first paper presents in detail data reduction and analysis techniques,
in addition to the results obtained for RXCJ\,0014.3-3022 ($z$=0.3068)
and RXCJ\,2308.3-0211 ($z$=0.2966).
It is organised as follows.
Section \ref{data} provides information on data quality and reduction,
whereas Sect.~\ref{results} describes the construction of the photometric
and spectroscopic catalogues, and the results obtained from
an analysis of the two cluster morphologies
and a spectral index analysis for a representative sample of cluster galaxies.
A comprehensive discussion follows (Sect.~\ref{discussion}),
whereas conclusions are summarised in Sect.~\ref{conclusions}.

Hereafter we adopt a $\mathrm{\Lambda}$CDM cosmology
where $\Omega_\mathrm{m} = 0.3$, $\Omega_\mathrm{\Lambda} = 0.7$,
and $h_{70} = H_0/70~\mathrm{km~s^{-1}~Mpc^{-1}} = 1$. At the average 
cluster redshift, 1 arcmin corresponds to 0.267 Mpc.


\section{Data: description and reduction}\label{data}

\subsection{X-ray imaging}\label{dataxray}

RXCJ\,0014-3022 and RXCJ\,2308.3-0211 were observed with XMM-\emph{Newton}
in AO-1 and AO-3 as part of the REFLEX-DXL cluster sample
(Zhang et al.~\cite{zhang06}).
Total exposure times were equal to 18.3 and 12 ks, respectively.
Observations were performed with thin filter for the three EPIC detectors:
\emph{MOS} data were collected in Full Frame (FF) mode,
while \emph{pn} data were taken in Extended Full Frame (EFF) in AO-1
and FF mode in AO-3, respectively.
For \emph{pn}, the fractions of the out-of-time (OOT) effect
are 2.32\% and 6.30\% for the EFF mode and FF mode, respectively.
An OOT event file is created and used to statistically remove the OOT effect.
We refer the reader to Zhang et al. (\cite{zhang06})
for the entire description of data reduction.
Here it is important to say that RXCJ\,0014-3022 and RXCJ\,2308.3-0211
have bolometric luminosities
equal to $2.12 (\pm 0.17) \times 10^{45}~\mathrm{erg~s^{-1}}$
and $1.20 (\pm 0.13) \times 10^{45}~\mathrm{erg~s^{-1}}$, respectively.
Furthermore, they were classified as offset centre and single objects,
respectively, according to the classification of the dynamical state
based on X-ray imaging (Jones \& Forman~\cite{jones92}).
This corresponds to the different nature of the two clusters,
one being a merging systems, the other a regular, centrally peaked,
cool-core cluster (Zhang et al.~\cite{zhang06} and references therein).

\subsection{Multi-object spectroscopy}\label{dataspec}

Spectroscopic observations in regions of the sky encompassing
RXCJ\,0014.3-3022 and RXCJ\,2308.3-0211 were performed
on August 14 through 16, 2004 (4.4 hr per cluster)
as part of the ESO Large Program 169.A-0595 (PI: H. B\"{o}hringer),
carried out in ESO GO time in visitor and service modes.
This program aims at maximising the number of member galaxies
lying in a wide region centred at the X-ray centroid of a cluster,
for 7 of the 13 DXL clusters.
This is achieved through multi-object spectroscopy (MOS)
in low resolution mode with VIMOS (VIsible Multi-Object Spectrograph),
which is mounted at the Nasmyth focus B of VLT-UT3 \emph{Melipal}
at Paranal Observatory (ESO), Chile.
VIMOS is a wide-field imager and multi-object spectrograph
operating in the visible (from 3600 to 10,000\,\AA).
It is made of four identical arms, each with a field of view (FOV)
of $7^{\prime} \times 8^{\prime}$ and a $0.205^{\prime \prime}$ pixel size,
separated by a gap between each quadrant of $\sim 2^{\prime}$.
Thus the total FOV is $4 \times 7^{\prime} \times 8^{\prime}$.
Each arm is equipped with 6 grisms, providing a spectral resolution
ranging from 200 (low) to 2500 (high), and one EEV CCD 4k $\times$ 2k.

Selection of targets for spectroscopy followed a strict luminosity criterion,
in order to minimise the probability of bias on future results
about cluster dynamics, galaxy populations, and star-formation activity,
which can potentially be introduced by additional colour selection.
I-band was chosen for pre-imaging, since it provides the closest analog
to a selection in stellar mass among the optical bands available for VIMOS.
In fact, I-band probes emission from stellar populations
mostly older than 1--2 Gyrs at z $\sim$ 0.3,
which are least affected by recent episodes of star-formation activity.
As a consequence, different environments are sampled with equal probability,
given the well-known existence of a correlation between density
and star-formation activity (Dressler~\cite{dressler80};
Whitmore et al.~\cite{whitmore93}).

\begin{figure}
   \centering
   \includegraphics[width=8 cm]{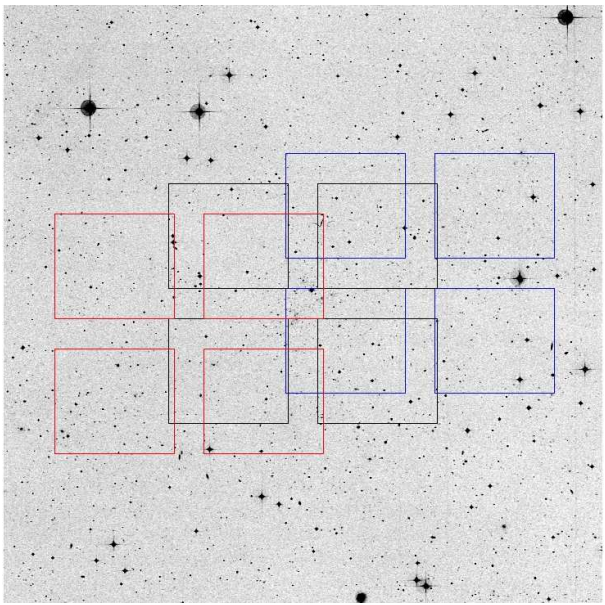}
      \caption{Mask setup for the VIMOS observations of RXCJ\,0014.3-3022, overlaid onto the corresponding DSS field. Each set of four boxes with the same colour represents one pointing; the VIMOS quadrants are aligned in the EW direction.}
         \label{vimosmasks}
   \end{figure}

Preimaging was executed adopting three different pointings with two ditherings of 15$^{\prime \prime}$ each, in order to allow the most complete coverage of the cluster central region, a good coverage of the cluster outskirts and sufficient overlap for the cross-shaped gaps between the CCDs. The three pointings are aligned in the E-W direction and partly overlapping, providing a total continuous coverage of about 34$^{\prime}$ in R.A. and 20 $^{\prime}$ in DEC. This corresponds to an uninterrupted coverage of the cluster out to 2.7$h_{70}^{-1}$ Mpc and a fairly good coverage of the outskirts out to 4.6 $h_{70}^{-1}$ Mpc at the clusters redshift (see Fig. \ref{vimosmasks}).

An 8-min exposure time per pointing, coupled to the photon collecting area
of the 8-m class VLT, enabled to completely sample galaxies
down to $I \sim 22.5$, which corresponds to about $M_\mathrm{I}^{\star}+2$
for the RXCJ\,0014.3-3022 core (see Couch et al.~\cite{couch98}),
where $M_\mathrm{I}^{\star}$ is the absolute magnitude
corresponding to the characteristic luminosity ($L^{\star}$)
of the I-band galaxy luminosity function
(LF, see Schechter~\cite{schechter76}).
Detection and identification of galaxies,
together with extraction of photometry, were accomplished
through {\it SExtractor} (Bertin \& Arnouts~\cite{bertin96}).
Galaxies with $17 \leq I \leq 19$ or $19 < I \leq 21.5$
were successively classified as bright or faint, respectively.
Two VIMOS masks per pointing were prepared using the VMMPS tool from ESO\footnote{http://www.eso.org/observing/p2pp/OSS/VMMPS}, one for the bright objects
and one for the faint ones.
The former were flagged as compulsory targets, whereas the latter were selected
as targets by the software, with the only requirement that the number
of objects per mask was maximised.
This approach allowed spectra to be obtained for most of the bright galaxies
whitout biasing the overall target selection; at the same time,
the LF of cluster galaxies was sampled across $L^{\star}$.

As for MOS observations, a minimum value of the signal-to-noise ratio (S/N)
equal to 10 (5) was required for bright (faint) galaxies,
which set the total exposure time equal to 6 (82) minutes.
This time was broken in three exposures per mask,
a total of six masks (three for bright objects and three for faint ones)
covering the region containing each cluster.
Mask were designed for the LR-Blue grism, which provides a spectral coverage
from 3700 to 6700\,\AA (observed frame) at a spectral resolution of about 200
for a $1^{\prime \prime}$ slitwidth, and does not suffer from fringing.
For objects at $z = 0.3$, the wavelength range 2830--5130\,\AA was mapped,
where important spectral features fall and were expected to be detected
on the basis of the S/N requirements.
Among them, the most relevant are the [OII], [OIII], H$_{\beta}$, H$_{\gamma}$,
and H$_{\delta}$ emission lines, the CaII$_\mathrm{H+K}$ absorption lines,
and the 4000 $\AA$ break.
These spectral features are extremely valuable
to determine spectroscopic redshifts up to $z \sim 0.8$
and increase their accuracy beyond that provided by the set-up.
Thanks to the multiplexing capability of the LR-Blue grism,
which enables to stack up to 4 spectra along the spectral dispersion direction,
up to 150--200 spectra per mask were collected, depending on brightness
and position of the targets.
Designed slits across individual targets included a region of empty sky
of $3^{\prime \prime}$ length, to provide a suitable sky subtraction.

Data from the MOS observations previously described were reduced
mainly using the dedicated software VIPGI\footnote{VIPGI (VIMOS Interactive Pipeline and Graphical Interface, Scodeggio et al.~\cite{scodeggio05}) is developed by the VIRMOS Consortium to handle the reduction of the VIMOS data for the VVDS (VIMOS VLT Deep Survey, Le F\`{e}vre et al.~\cite{lefevre05})},
which is a complete data reduction environment specifically developed
to handle and reduce data from imaging, MOS and integral-field-unit
spectroscopy with VIMOS.
Its capabilities and quality have been tested against the reduction
of many tens of thousands spectra from the VVDS, and are thoroughly explained
in the aforementioned publications.
In addition to bias and flatfield correction, multiplex spectra identification,
sky subtraction, spectral reduction, and wavelength calibration,
VIPGI is capable of data editing and redshift evaluation
based on single- and multiple-line fitting, after manual selection
of candidate lines to be matched with available line catalogues.
Data reduction followed the standard approach,
fully described in the VIPGI manual
and in Scodeggio et al. (\cite{scodeggio05}).
A master bias frame was obtained, where possible,
by combining all the available bias frames from the night;
rejection of overscan areas and/or bad pixels was performed
by standard $3 \sigma$ clipping through the data cube
defined by the bias frames.
In particular, the adjustment of initial guesses for wavelength calibration
was usually fairly quick: a check of the line catalogue of the arcs
always produced a good agreement with the positions of features
in the arc spectra, so only small shifts of 1 to 3 pixels
(i.e., about 4 to 15\,\AA at the resolution of the LR-Blue grism)
were introduced as a correction.
None of the quadrants was found to be tilted or rotated.
The complete adjustment of the calibration arcs
was usually performed with a small shift of less than 5 pixels
(i.e., about 20\,\AA) on the dispersion direction.
Spectra detection, based on the adjusted arc calibration,
was performed using standard parameters too.
After manual correction of shifted lines slit by slit,
the wavelength calibration shows a typical rms of $\sim 1.5$\,\AA.

As for science spectra, a combined reduction of the three frames per mask
was performed with the \emph{Reduce Sequence of Observations} tool of VIPGI.
This avoids confusion with noise and residuals from non optimal bias removal
and flat-field correction; in addition, it enables detection of objects
that are fainter than expected.
The minimum detection level was set equal to $3 \sigma$
for bright and faint targets, where sky subtraction was based
on a fit with a 3rd-order polynomial over each single frame
instead of using a simple median over the stack of three frames.
This procedure was found to provide a better estimation of the sky emission
and more robust detections: using a median-based sky subtraction instead,
up to 15\% of the faint targets could remain undetected.
Spectra of individual targets were extracted using Horne's algorithm
(Horne~\cite{horne86}); standard parameters for detection and deblending
(i.e., disentangling of multiple spectra falling in the same slit) were used,
as they provided a robust extraction.
At the end, all spectra were inspected by eye to check
if residual sky contamination, hot pixels,
ghosts arising from the superposition of different spectral orders,
and other artifacts were present,
and if all targets had been successfully extracted.
In addition, initial values of the S/N were computed
for the continuum emission and spectral features.

Spectroscopic redshifts were mainly determined through VIPGI
(the EZ\footnote{http://cosmos.iasf-milano.inaf.it/pandora/} package,
plus software for the manual detection and fit of spectral features).
After a pipeline determination for all spectra,
individual redshifts were inspected by eye
and manual determination was executed in case of a probable misidentifying.
In general, spectra with initial values of the S/N equal to 4 (at least)
for the continuum and larger than 10 for spectral features
enabled a clear determination of spectral types,
and, thus, provided an unambiguous determination of redshifts.
Typically a reliable spectrum could not be extracted
from data with a mean S/N less than 3;
the corresponding slits were flagged for removal.
Non-detections were used to evaluate the mean spectral noise,
which amounts to $\sim 20$ counts/pixel, the blue spectral region
covered by the LR-Blue grism (i.e., $\lambda < 5000$\,\AA)
being slightly noisier than the red one.
The faintest, reliably detected spectra have continuum fluxes
from 100 to 500 counts/pixel, depending on the spectral type,
and spectral features with high S/N.

%

   \begin{figure}
   \centering
   \includegraphics[width=8 cm]{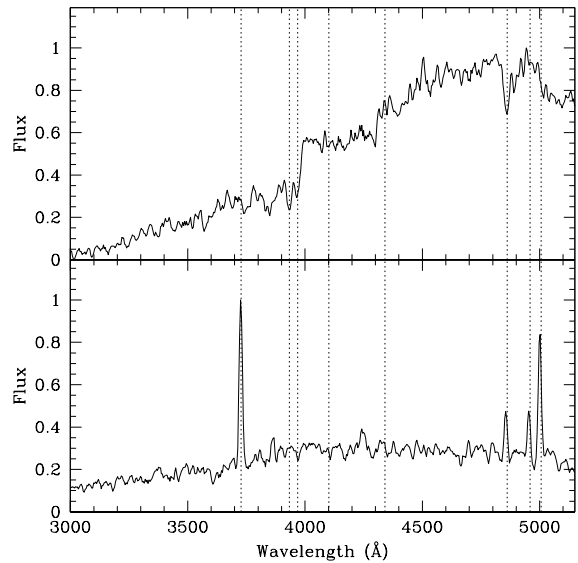}
      \caption{Examples of VIMOS spectra (rest frame, normalised) for two members of RXCJ\,2308.3-0211, with typical spectral features marked. Top: an elliptical galaxy, with Ca\,II$_\mathrm{H+K}$($\lambda \lambda$ 3933.7, 3968.5\,\AA) absorption lines and H$_\mathrm{\beta}$($\lambda$ 4861.3\,\AA) feature in absorption. Bottom: an emission-line galaxy, with [O\,II]($\lambda$ 3727.5\,\AA), H$_\mathrm{\beta}$, and [O\,III]($\lambda \lambda$ 4958.9, 5006.8\,\AA) emission features.}
         \label{vimosspectra}
   \end{figure}

\subsection{Wide-field imaging}\label{dataphot}

Optical photometry was carried out using the wide-field imager (WFI)
mounted at the Cassegrain focus of ESO/MPG-2.2m telescope at La Silla, Chile.
WFI (Baade et al.~\cite{baade99}) is a focal reducer-type mosaic camera
which consists of $ 4 \times 2$ CCD chips, each with $2048 \times 4096$ pixels
and a FOV of $8.12^{\prime} \times 16.25^{\prime}$
($0.238^{\prime \prime}$/pixel).
Chips are separated by gaps of $23.8^{\prime \prime}$
and $14.3^{\prime \prime}$ in the R.A. and Dec directions, respectively,
so the WFI FOV is $34^{\prime} \times 33^{\prime}$,
with a filling factor of 95.9\%.
The data presented here were obtained as part of a program
carried out in visitor and service modes (P.I.: H. B{\"o}hringer)
during MPG observing time.
In particular, observations of RXCJ\,0014.3-3022 and RXCJ\,2308.3-0211
in the B, V, and R passbands were performed on September 27 through 30, 2000,
and in photometric conditions.
They were split into sequences of eight dithered sub-exposures.
Total exposure times are listed in Table~1.
Filter curves can be found in Arnouts et al. (\cite{arnouts01})
and on the web page of the la Silla Science Operations Team\footnote{http://www.ls.eso.org/lasilla/sciops/2p2/E2p2M/WFI/filters}.
Standard stars were observed in all the four nights:
three Landolt fields (Landolt~\cite{landolt92}) were targeted
for a total of 16 standard star OBs per filter.

The WFI data were reduced using the data reduction system
developed for the ESO Imaging Survey
(EIS, Renzini \& da Costa~\cite{renzini97})
and its associated EIS/MVM image processing library version 1.0.1
({\it Alambic}, Vandame 2004).
{\it Alambic} is a publicly available\footnote{http://www.eso.org/science/eis/survey\_release.html} software designed to automatically transform raw images
from single/multi-chip optical/infrared cameras into reduced images
for scientific use.
In addition to the standard bias-subtraction, flatfield correction,
and trimming, the EIS/MVM image processing pipeline performs
background estimation, de-fringing (if needed), astrometric calibration,
minimisation of chip-to-chip variations in sensitivity,
and detection/masking of satellite tracks.
All these steps are described in detail in previous publications
(Arnouts et al.~\cite{arnouts01}; Vandame~\cite{vandame04};
Mignano et al.~\cite{mignano07}; Pierini et al.~\cite{pierini08}),
and, thus, are skipped here.
However, we note that the astrometric calibration was derived
using the GSC2.2 reference catalogue and a distortion model
described by a second order polynomial.
Its precision is $0.2^{\prime \prime}$ but the internal accuracy
is about 70\,mas (see Mignano et al.~\cite{mignano07}).
Furthermore, we note that the CCD-to-CCD gain variations were corrected
using median background values sampled in sub-regions bordering adjacent CCDs,
but no illumination correction was applied.
This may lead to relative zeropoint offsets from the centre to the borders
of the image of up to 10\%, according to the {\it still experimental} method
to derive a zeropoint correction map designed by Selman (\cite{selman04}).

Source detection and photometry were based on SExtractor
(Bertin \& Arnouts~\cite{bertin96}) both for standard and science images.
Magnitudes were calibrated to the Johnson-Cousins system
using Landolt (\cite{landolt92}) standard stars
whose magnitudes were obtained using a 10~arcsec-wide circular aperture,
which proved to be adequate by monitoring the growth curve
of all the measured stars.
Photometric standards were observed over a rather broad range of airmasses,
but science frames were taken at the best airmass per target;
so the photometric solutions used to calibrate reduced images
were obtained from merging all the measurements of standard stars
for each passband.
The number of non-saturated Landolt stars per field did not allow
independent solutions to be determined for each of the eight chips of WFI.
Hence calibration had to rely on solutions based on measurements
taken across all chips.
Despite the fact that the EIS data reduction system
includes a photometric pipeline for the automatic determination
of photometric solutions, these were determined interactively
using the {\it IRAF}\footnote{{\it IRAF} is the Image Reduction and Analysis Facility, a general purpose software system for the reduction and analysis of astronomical data. IRAF is written and supported by the IRAF programming group at the National Optical Astronomy Observatories (NOAO) in Tucson, Arizona. NOAO is operated by the Association of Universities for Research in Astronomy (AURA), Inc. under cooperative agreement with the National Science Foundation.} task {\it fitparams}.
This choice allows interactive rejection of individual measurements, stars,
and chips.
Photometric solutions with minimum scatter were obtained
through a two-parameter linear fit to about 200 photometric points
per passband, the extinction coefficient being set equal
to that listed in the ``definitive'' solution
obtained by the 2p2 Telescope Team\footnote{http://www.ls.eso.org/lasilla/Telescopes/2p2T/E2p2M/WFI/zeropoints/}.
In general, zeropoints and colour terms are consistent with those
obtained by the 2p2 Telescope Team or by the ESO DEEP Public Survey (DPS) team
(Mignano et al.~\cite{mignano07}), as shown by the comparison
of Tables~2, 3, and 4.

As for science images, source extraction and photometry were obtained
after matching the BVR images of each target to the worst seeing
(1 and $1.3^{\prime \prime}$ FWHM for RXCJ\,0014.3-3022
and RXCJ\,2308.3-0211, respectively), using the {\it IRAF} task {\it psfmatch},
and taking into account the weight-maps associated with the individual images,
produced by {\it Alambic}.
A common configuration file was used to produce three catalogues per target,
a minimum number of trivial adjustments (e.g., seeing, zeropoint)
being made for individual images.
The deepest R-band image was used as the detection image,
where sources are defined by an area with a minimum number of 5 pixels
above a threshold of $1~\sigma$ of the background counts.
Source photometry in individual passbands was extracted
in fixed circular apertures (from $1.2^{\prime \prime}$
to $10^{\prime \prime}$ in diameter) or in flexible elliptical apertures
(Kron-like, Kron~\cite{kron80}) with a Kron-factor of 2.5
and a minim radius of 3.5 pixels.
Total magnitudes computed from the latter photometry (Kron-like)
are used hereafter.
Background subtraction was based on values taken directly
from the background map, as obtained from a bi-cubic-spline interpolation,
after applying median-filtering to sub-regions of an image
equivalent to $3 \times 3$ meshes, the mesh-size being equal to 64 pixels
(i.e., $15.2^{\prime \prime}$).
Object magnitudes were corrected for galactic extinction
according to the Schlegel et al. (\cite{schlegel98}) Galactic reddening maps
(from {\it NED}) and converted to the AB system
according to the response function of the optical system
(see Alcal\'a et al.~\cite{alcala04}).
The output catalogues were successively pruned for fake sources by hand
before photometric redshifts were determined.
In addition, stars and galaxies could be safely identified
on the basis of their surface brightness profile and optical colours
down to $\mathrm{R}=21.5$.
Fainter than this limit, number counts are dominated by galaxies,
so that all detected objects with $\mathrm{R}>21.5$
are assumed to be ``bona fide'' galaxies.

\begin{figure}
   \centering
   \includegraphics[width=8 cm]{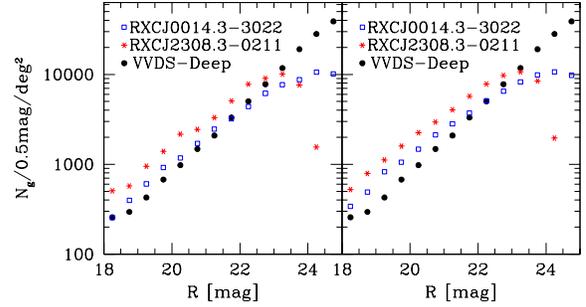}
      \caption{Comparison of number counts for both clusters and the VVDS-Deep (McCracken et al.~\cite{mccracken03}). Left: number counts in a neighbouring, low density region of each cluster. Right: number counts for the entire region of a cluster imaged with WFI.}
         \label{ncounts}
   \end{figure}

Depth and quality of the final catalogues were assessed in two ways.
Firstly, we compared the number counts of stars safely identified
in the R-band image of each cluster with those expected from the model
of stellar population synthesis of the Galaxy by Robin et al. (\cite{robin03})
for $\mathrm{R} \le 21.5$.
A good agreement between measured and expected values was found
within the uncertainties.
Secondly, we compared the number counts of galaxies,
falling either within the entire mapped region of a cluster
or in a low-density subregion (i.e., the ``field'', see Sect. \ref{clusmorph}),
with the galaxy number counts determined in deep fields
(in particular VVDS Deep, McCracken et al.~\cite{mccracken03}).
In general, our galaxy number counts exceed those obtained
from observations of deep fields for $\mathrm{R} \le 21.5$,
where our galaxies are safely identified (cf. Fig.~\ref{ncounts}).
Unsurprisingly, this is especially true when the entire mapped regions
of RXCJ\,0014.3-3022 and RXCJ\,2308.3-0211 are considered.
On the other hand, the number counts of all galaxies,
falling either in the ``field'' or in the entire mapped region of a cluster,
start falling short of the galaxy number counts in deep fields
at $\mathrm{R}=22$--23.
In particular, this deficit amounts to a factor of 2 at around 23.5\,R-mag,
where the number of background galaxies dominates the number
of likely cluster members.
Assuming as a completeness limit the magnitude at which the observed counts
equal to 50\% of the expected ones, we thus conclude that our R-band selected catalogues
are complete down to $\sim 23.5~\mathrm{mag}$.

This holds in spite of the different decrease of number counts
in the two cluster regions (see Fig.~\ref{ncounts}).
The different behaviour at $\mathrm{R} > 23.5~\mathrm{mag}$
must be ascribed to a combination of Galactic absorption and seeing.
In fact, RXCJ\,0014.3-3022 lies in a sky region far away
from the Galactic Plane and was observed at a typical seeing
of $\sim 1^{\prime \prime}$; conversely, RXCJ\,2308.3-0211 lies
close to the Galactic Plane and was observed at a typical seeing
of $\sim 1.2^{\prime \prime}$.
It is then likely that a larger fraction of objects fainter
than $\sim 23.5~\mathrm{mag}$ has failed detection in the second region.
Consistently, catalogues exhibit comparable numbers
of objects with $\mathrm{R} \le 23.5~\mathrm{mag}$: 
14,409 for RXCJ\,0014.3-3022 and 14,820 for RXCJ\,2308.3-0211.
A Kolmogorov-Smirnov test confirms that number counts
for the two complete samples are consistent:
the probability of a difference is equal to only 7.1\%.

   \begin{table}
      \caption[]{The  photometric solutions available in this work. The table lists: in Col. 1 the passband; in Cols. 2 -- 4 the zeropoint in the Vega magnitude system (ZP), the extinction coefficient (k), and the colour term (CT) together with their errors. These ``best fit'' parameters were obtained from a two-parameter fit to about 200 measurements across the WFI field for each passband, the extinction coefficient being fixed.}
         \label{photour}
     $$
         \begin{array}{cccc}
            \hline
            \noalign{\smallskip}
            \mathrm{Passband} & ZP & k & CT \\
            \noalign{\smallskip}
            \hline
            \noalign{\smallskip}
            B & 24.66 \pm 0.005 & 0.22 & 0.26 \pm 0.007 \\
            V & 24.23 \pm 0.006 & 0.11 & -0.15 \pm 0.012 \\
            R & 24.50 \pm 0.007 & 0.07 & -0.02 \pm 0.015 \\
            \noalign{\smallskip}
            \hline
         \end{array}
     $$
   \end{table}
%

   \begin{table}
      \caption[]{The ``definitive'' photometric solutions obtained by the 2p2 Telescope Team from observations of standard stars in perfectly photometric nights, where a bunch of standard fields were moved around each chip of WFI. All parameters were fitted simultaneously as free parameters, with good airmass and colour range, and around 300 stars per fit. The table below gives the average solutions over all chips.}
         \label{phot2p2}
     $$
         \begin{array}{cccc}
            \hline
            \noalign{\smallskip}
            \mathrm{Passband} & ZP & k & CT \\
            \noalign{\smallskip}
            \hline
            \noalign{\smallskip}
            B & 24.81 \pm 0.05 & 0.22 \pm 0.015 & 0.25 \pm 0.01 \\
            V & 24.15 \pm 0.04 & 0.11 \pm 0.01 & -0.13 \pm 0.01 \\
            R & 24.47 \pm 0.04 & 0.07 \pm 0.01 & 0.00 \pm 0.00 \\
            \noalign{\smallskip}
            \hline
         \end{array}
     $$
   \end{table}
%

   \begin{table}
      \caption[]{The table below gives median values for all the photometric solutions based on three-parameter fits obtained by the ESO DPS team.}
         \label{photdps}
     $$
         \begin{array}{cccc}
            \hline
            \noalign{\smallskip}
            \mathrm{Passband} & ZP & k & CT \\
            \noalign{\smallskip}
            \hline
            \noalign{\smallskip}
            B & 24.58 & 0.22 & 0.24 \\
            V & 24.23 & 0.19 & -0.04 \\
            R & 24.49 & 0.08 & -0.01 \\
            \noalign{\smallskip}
            \hline
         \end{array}
     $$
   \end{table}
%

\subsection{GALEX UV photometry}\label{datauv}

In order to better constrain the recent star-formation activity
in member galaxies of RXCJ\,0014.3-3022 and RXCJ\,2308.3-0211,
we complemented the optical data with GALEX UV photometry
The GALEX data archive was scanned via the MultiMission Archive
at Space Telescope Science Institute (MAST\footnote{http://galex.stsci.edu}):
positions and total magnitudes of galaxies falling in the regions
of both clusters were extracted from the All Sky Imaging Survey
(GR3 data release).
The total UV magnitude of an object was calculated as a weighted mean
when multiple detections of the same object exist
from overlapping GALEX pointings. 
In general, cross-identification of GALEX sources
with objects in our catalogues was largely unambiguous.
When multiple catalogue objects were present within the radius
of the GALEX PSF ($6^{\prime \prime}$ FWHM), the bluest one
(i.e., with smaller (B-R) colour) was identified as the counterpart
of the UV source.
In the absence of GALEX FUV detections, $3 \sigma$ upper limits
on the UV magnitude are computed from the completeness FUV magnitude
of the two fields imaged by GALEX.

\section{Results}\label{results}

\subsection{Photometric redshifts}\label{catphot}

Cluster member galaxies are identified
on the basis of photometric or spectroscopic redshifts;
two catalogues are built accordingly.
In the first case, member galaxies have photo-$z$'s
consistent with the photo-$z$ of the cluster;
hence they are``bona fide'' members.
In this section, it is described how we determine photometric redshifts,
estimate their accuracy, and establish cluster membership based on photo-$z$'s.

\begin{figure}
   \centering
   \includegraphics[width=4 cm]{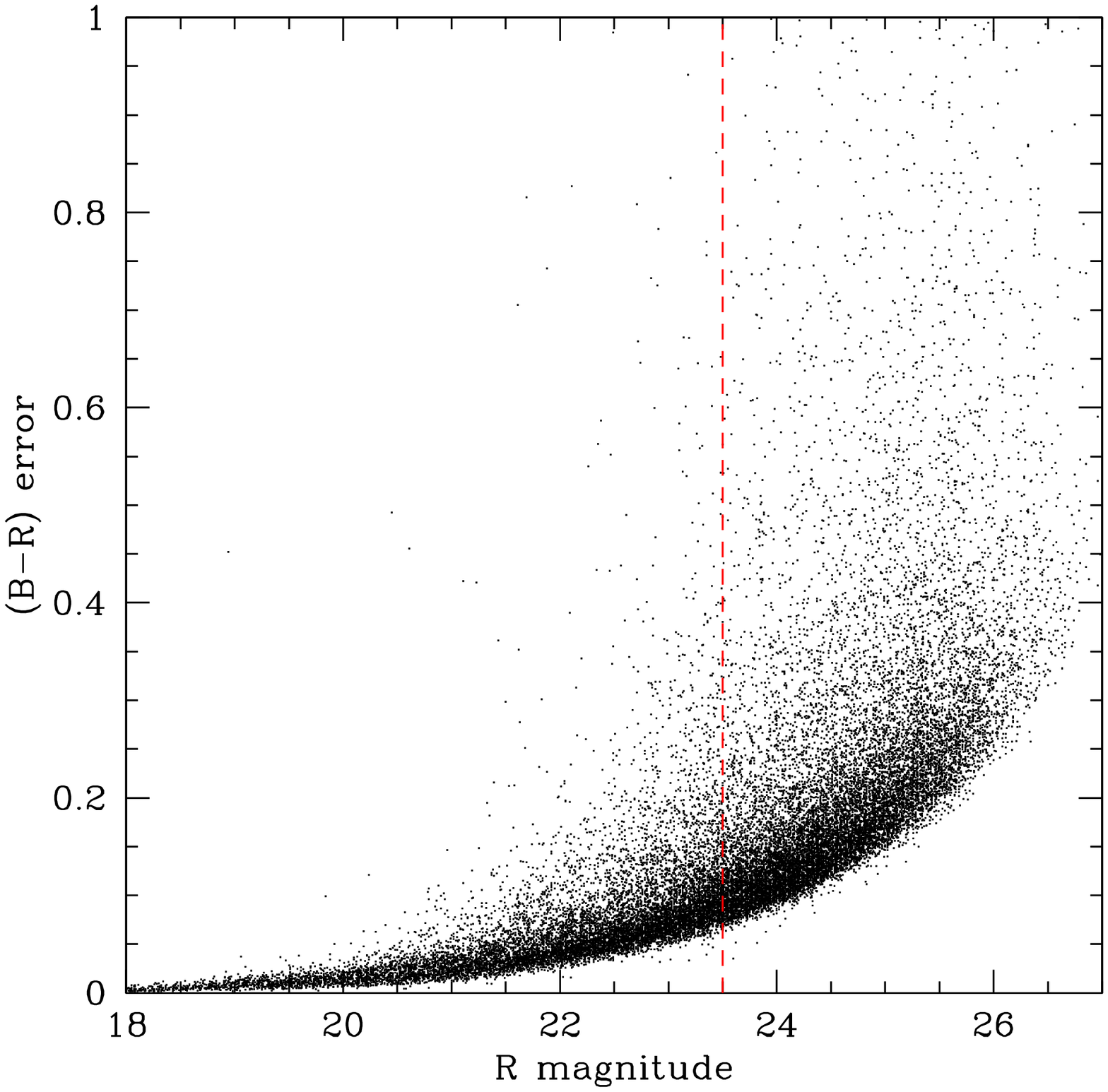}
   \includegraphics[width=4 cm]{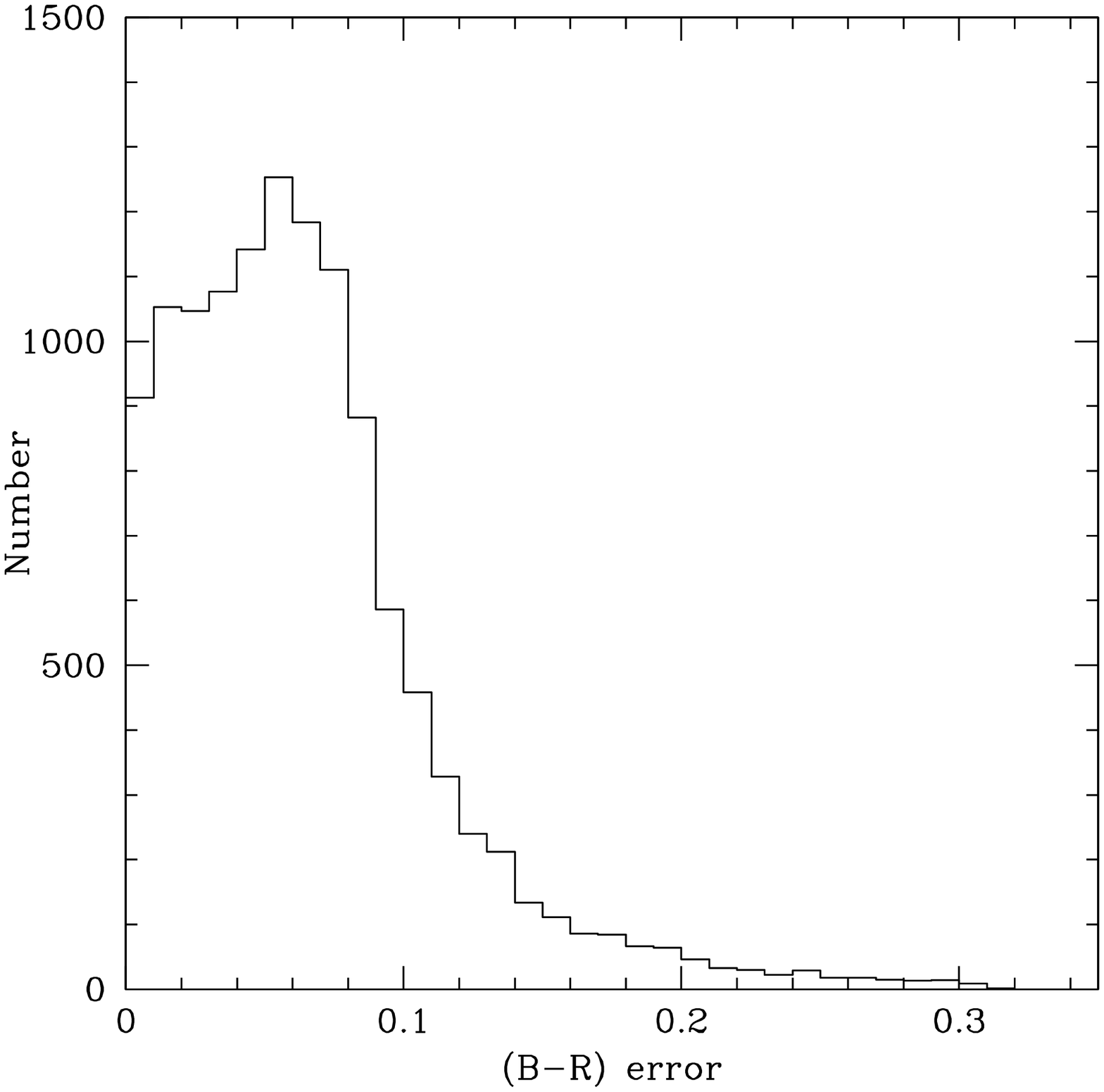}
      \caption{Photometric quality of WFI data for RXCJ\,0014.3-3022. Left: error on the $(B-R)$ colour as a function of R-band magnitude ($R$). A vertical dashed line marks the completeness limit $R = 23.5$. Right: distribution of errors on $(B-R)$ for the complete sample.}
         \label{magerrs}
   \end{figure}

Photometric redshifts are obtained by using a prior based
on the $(B-R)$ colour\footnote{At the redshifts of the two clusters, the B and R broad-band filters map, respectively, the rest-frame wavelength ranges 3022--3940\,\AA and 4247--6045\,\AA. Hence, these two passbands bracket the 4000\,\AA break. In addition, B band contains the redshifted emission in the [OII] line.}, given the limited photometry available.
This information is accurate enough only for objects
brighter than the completeness limit of the photometric catalogues
pruned for fake objects and stars brighter than $\mathrm{R}=21.5$
(see Sect.~\ref{dataphot}).
This is illustrated by the distribution of the errors on $(B-R)$
as a function of R-band magnitude for the 23,900 objects
in the region of RXCJ\,0014.3-3022 with $14 < R < 27$
(see Fig.~\ref{magerrs}).
Objects brighter than $R \sim 23.5$ exhibit a typical error
on $(B-R)$ of about 0.05 mag; only 248 objects among them
(i.e., 0.02\% of the complete sample) exhibit errors larger than 0.2 mag.
The latter have $21.4 < R \le 23.5$, with a mean $R = 23.05$.

\begin{figure}
   \centering
   \includegraphics[width=8 cm]{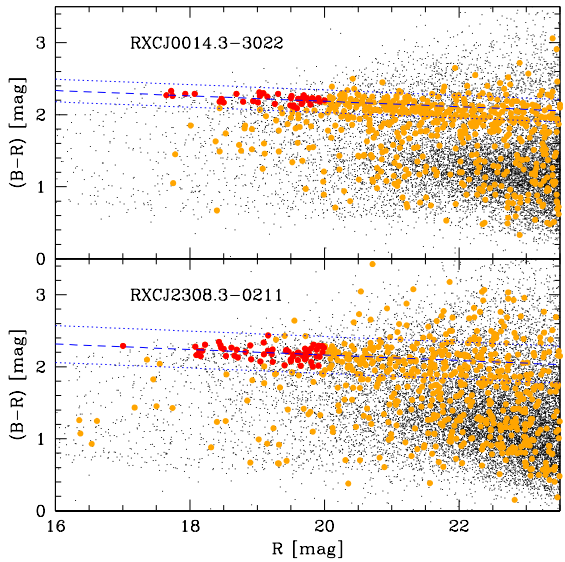}
      \caption{$(B-R)$ vs. $R$ colour--magnitude diagram (observed frame) for the regions of RXCJ\,0014-3022 (upper panel) and RXCJ\,2308.3-0211 (lower panel) imaged with WFI. Orange points mark galaxies within $3^{\prime}$ from the cluster centre: most of them define a red-sequence. Red circles mark galaxies used for fitting the red sequence; this locus is reproduced (dashed line) together with the $\pm 3 \sigma$ scatter around it (dotted lines).}
         \label{rseqfield}
   \end{figure}

$(B-R)$ vs. $R$ colour--magnitude (CM) diagrams are successively built
from the complete catalogues of the two clusters,
after removing some galaxies with photometry compromised
by intervening bright stars.
This action does not impact any of the following results.
A visual inspection of both CM diagrams (see Fig.~\ref{rseqfield})
clearly reveals the presence of a red sequence,
which exhibits $(B-R) \sim 2.2$ (observed frame).
In each case, the red sequence was fitted
using an iterative $3 \sigma$ clipping linear regression algorithm,
all objects lying within a distance of $3^{\prime}$
from X-ray centroid of the cluster and with $17 \leq R \leq 20$
being considered in the first step.
The best-fits of the two red sequences are remarkably coincident,
as shown later on.
Thus objects in the complete samples of the two clusters
are classified as ``red'' if they lie within $3 \sigma$
from the best-fit equation of the red sequence,
or ``blue'' if their $(B-R)$ colours are bluer than this locus
at a significance higher than $3 \sigma$ (see Fig.~\ref{rseqfield}).

This separation provides some guidance for selecting the SED templates
that are explored to determine the photometric redshift of each object
through the well-tested code {\sl HyperZ}
(Bolzonella et al.~\cite{bolzonella00}).
Specifically, red galaxies were only fitted using a suite of templates
for galaxies dominated by old, passively-evolving stellar populations
(E and S0 types), whereas blue objects were fitted with a suite of templates
for star-forming galaxies (spirals and irregulars).
Photometric redshifts were also determined without any prior on colour
(i.e., morphology), which gave consistent results.
Default synthetic non-evolving templates for E, S0, Sa, Sb, Sc, Sd,
and Im galaxies were found to give the best results.
Best-fit solutions from {\sl HyperZ} were trained on spectroscopic redshifts,
which provided an rms uncertainty of 0.06 at $z=0.3$ (see BPB07).

The photometric redshift of a cluster is defined
as the mean of the photo-$z$'s of its spectroscopic members,
as identified in Sect.~\ref{catspec}.
All galaxies with photo-$z$'s within $\pm 1 \sigma$ (i.e., $\pm 0.06$)
from $z_\mathrm{phot}$ are defined as cluster photometric members.

%

%

\subsubsection{RXCJ\,0014.3-3022: photometric members}\label{a2744photmem}

The CM diagram of the inner region of RXCJ\,0014.3-3022 shows that
five galaxies populate a narrow bin ($17.6 \le R \le 18.1$)
at the bright-end of the red sequence.
Four of them are part of two visual pairs of BCGs
associated with the two X-ray subcomponents of the cluster
(see Pierini et al.~\cite{pierini08}).

Fitting the CM distribution of 41 red-sequence objects,
selected with a recursive algorithm, gives:
\begin{displaymath}
(B-R) = (2.935\pm0.238) - (0.037\pm0.003) \times R.
\end{displaymath}
The rms scatter around this locus is equal to 0.052.

RXCJ\,0014.3-3022 has $z_\mathrm{phot}=0.34$
and its bona fide members have a photometric redshift between 0.28 and 0.4.
Its photometric redshift is thus higher than the spectroscopic one
($z=0.3068$), although the two redshifts are consistent
given the large typical uncertainty of photo-$z$'s.
The discrepancy is not driven by contamination from outliers,
which is below 10\%, or by the overall robustness of the identification
of cluster members (see BPB07), but by blue members,
which exhibit higher photo-$z$'s and a larger scatter
with respect to the red members, on average.
This is mostly due to the presence of large peculiar velocities
(up to $4000~\mathrm{km~s^{-1}}$, see BPB07),
which can produce a difference in $z$ of up to $\sim 0.02$.
The coupling of a relatively large, positive excess in the velocity
along the line of sight and the rest-frame wavelength coverage
offered by the filter set can boost photo-$z$'s by about 0.04,
but only for blue cluster member galaxies.
In fact, at the cluster spectroscopic redshift,
the redshifted [OII] emission line falls in the red tail of the B filter,
and just shortwards of the blue edge of the V passband,
whereas the 4000\,\AA~break falls in between the two passbands.
The former feature characterises star-forming systems,
the latter is stronger in old, passively evolving galaxies.
Consistently, the photo-$z$'s of red galaxies ($\sim 0.32$)
better agree with the spectroscopic redshift of the cluster.

In conclusion, the catalogue of photometric members
of RXCJ\,0014.3-3022 comprises 3698 galaxies
across a 900 square arcmin-region centred at the X-ray centroid of the cluster,
of which 1409 (2289) are classified as red (blue).

\subsubsection{RXCJ\,2308.3-0211: photometric members}\label{a2537photmem}

The inner region of RXCJ\,2308.3-0211 exhibits a similar red-sequence
to that of RXCJ\,0014.3-3022, but also three interesting differences.
First, RXCJ\,2308.3-0211 hosts a single BCG,
which is brighter than other red photometric members
by at least $\sim 1$~R-mag.
Second, its red-sequence is more populated
(64 photometric members against 41) and exhibits a 60\% larger RMS scatter
(0.085 against 0.052).
Third, red members amount to 57\% of the total in RXCJ\,2308.3-0211
against a value of 38\% in RXCJ\,0014.3-3022.

A fit to the red sequence defined by 64 robustly selected objects gives:
\begin{displaymath}
(B-R) = (2.933\pm0.323) - (0.038\pm0.003) \times R
\end{displaymath}
with a scatter of 0.085.
Photometric ($z_\mathrm{phot}=0.29$) and spectroscopic ($z=0.2966$) redshifts
are in excellent agreement for this cluster: thus, galaxies with photo-$z$'s
between 0.23 and 0.35 are identified as members.
The catalogue of photometric members of RXCJ\,2308.3-0211
contains 4109 objects across a 900 square arcmin-region
centred at the X-ray centroid of the cluster,
of which 2361 (1748) are classified as red (blue).

\subsection{Spectroscopic redshifts}\label{catspec}

Spectroscopically confirmed members of either cluster are identified
from the available spectroscopic catalogues as follows.
As a first step, a relatively broad distribution in spectroscopic redshifts
is considered, by selecting a range centred on the mean cluster redshift
and with a width of 0.05.
This is a fairly large value: it corresponds to a velocity range
of about 10,000 $\mathrm{km~s^{-1}}$ at $z \sim 0.3$.
However, it avoids the ``a priori'' loss of high-speed cluster members
and still enables unambiguous rejection of any other structure
present in the foreground/background.

After this selection, cluster membership is estimated
by using the biweight estimators introduced by Beers et al. (\cite{beers90})
and successively applied with success in many different studies
(e.g. Biviano et al.~\cite{biviano92}; Girardi et al.~\cite{girardi93};
Adami et al.~\cite{adami98}; Girardi et al.~\cite{girardi98};
Biviano et al.~\cite{biviano06}).
Successively, all objects with peculiar velocities
larger than $\pm 4000~\mathrm{km~s^{-1}}$ are rejected.
Interlopers are then removed following the algorithm of den Hartog \& Katgert
(\cite{denhartog96}, see also Katgert et al.~\cite{katgert04}),
which identifies galaxies unlikely to be bound to the cluster on the basis of their location in projected phase-space. Velocity errors and relativistic correction are applied to the cluster
velocity dispersion estimate, as explained in Danese, De Zotti \& di Tullio (\cite{danese80}).
The X-ray centroid of each cluster was used as reference in these calculations.
The application of the overall procedure is discussed
in Braglia et al. (2009b, in prep.),
whereas the method is validated by the analysis of clusters extracted from cosmological numerical simulations in Biviano et al. (\cite{biviano06}).

Finally, each catalogue of spectroscopic cluster members
is cross-correlated with the photometric one, which gives a new catalogue
combining photometric and spectroscopic information.
All detected spectroscopic targets are brighter than 23.5 R-mag, as expected.
For this cross-correlation, the celestial coordinates
of spectroscopic members were taken from pre-imaging.
No correction for seeing was introduced since pre-imaging
and R-band imaging were executed in similar seeing conditions.
The catalogues of spectroscopic members for the two clusters
are provided in Appendices B and C; they contain photometric information
from WFI imaging as well as the spectral indices
computed in Sect.~\ref{indices}.

\subsubsection{RXCJ\,0014.3-3022: spectroscopic members}\label{a2744specmem}

\begin{figure}
   \centering
   \includegraphics[width=8 cm]{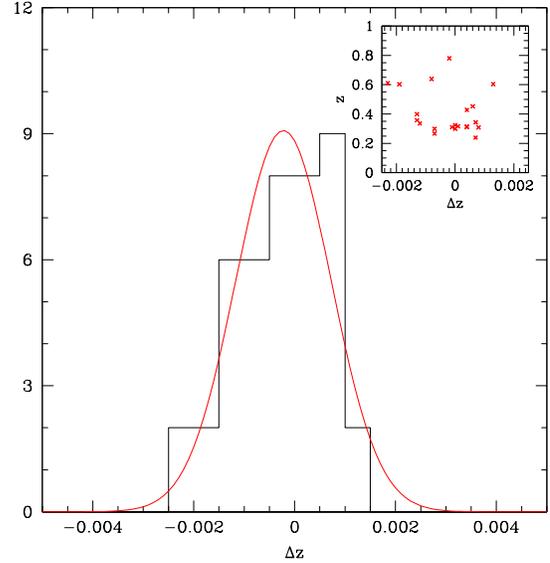}
      \caption{Distribution of the errors on spectroscopic redshifts for multiply-observed galaxies in the region of RXCJ\,0014.3-3022. IN the main plot, a gaussian fit to this distribution is also reproduced. In the upper right inset, the same errors (in abscissa) are plotted against the associated redshifts. No clear correlation is found for these galaxies, which mostly lie at the same redshift of the cluster.}
         \label{zerrors}
   \end{figure}

The spectroscopic data reduction produced a total of 545 slits
with at least one object in the entire region of RXCJ\,0014.3-3022
covered with VIMOS.
This corresponds to a total of 872 objects,
since 147 slits contained multiple objects (up to 5).
Among them, there are 64 serendipitous faint stars.
For 262 objects, the S/N is too low for the extraction of a reliable spectrum,
so they were flagged as non-detections. 
Out of the remaining 546 detected non-stellar objects,
43 were found to be multiple observations of the same galaxies.
From these repeated observations, we estimate a mean velocity error
equal to $294 \pm 50~\mathrm{km~s^{-1}}$, which is in good agreement
with the value of $276~\mathrm{km~s^{-1}}$ obtained for the VVDS 
using the same VIMOS instrumental setup.
The error distribution for the corresponding pairs of spectroscopic redshift
is plotted in Fig.~\ref{zerrors}.
In conclusion, reliable VIMOS spectra are available
for 512 non-stellar objects, which span the I-band magnitude range
from 19 to about 22.5, and reach $z=0.85$\footnote{Three quasars at $z=1.02$, 2.51, and 3.08 were also detected.}.

This sample was expanded by considering 101 additional objects in the field of RXCJ\,0014.3-3022,
selected from publicly available spectroscopic redshifts in \emph{NED}.
Cross-correlation of this new sample with our WFI photometric catalogue
showed that the additional objects are mainly brighter than $R=22.0$
(see Fig. \ref{rseqspec}).
Furthermore, 24 out of the 101 objects with spectroscopic redshifts
in \emph{NED} corresponded to objects at $0.06 \le z \le 0.35$
and observed by us.
Comparison of the photometric redshifts of the corresponding pairs
allowed the external error of our redshift determinations to be assessed:
it is equal to $276 \pm 55~\mathrm{km~s^{-1}}$, on average,
which is fully consistent with the internal error of the same measurements.

The final catalogue of spectroscopically confirmed members of RXCJ\,0014.3-3022
comprises 190 different objects, 101 of which have VIMOS spectra.
In the corresponding CM diagram, the red-sequence is visible
(see Fig.~\ref{rseqspec}); its fit, according to the same selection criteria
as those used in Sect.~\ref{a2744photmem}, is:
\begin{displaymath}
(B-R) = (3.023\pm0.277) -(0.042\pm0.004)\times R.
\end{displaymath}
The red sequence defined by spectroscopic members
lying within $3^{\prime}$ from the centre of RXCJ\,0014.3-3022
and with $17 \le R \le 20$ exhibits a slightly higher
normalisation and RMS scatter (0.049), together with a slightly steeper slope,
than the locus defined by analogous photometric cluster members.
Differences are not statistically significant however.

Finally, we note that there are three very blue objects
(i.e., with $(B-R) \leq 1$) even in the inner region of RXCJ\,0014.3-3022,
i.e., within a cluster-centric distance of about 800 kpc.
This is consistent with the presence of enhanced star-formation activity
even there, as found by Couch et al. (\cite{couch98}) and BPB07,
although phase-space projection effects cannot be completely ruled out.

\subsubsection{RXCJ\,2308.3-0211: spectroscopic members}\label{a2537specmem}

For the region encompassing RXCJ\,2308.3-0211, a total of 1122 slits
is available from our VIMOS spectroscopy.
They provide 1522 spectra, since 424 slits contained multiple objects:
64 spectra correspond to faint stars, whereas 446 have a too poor S/N.
As a result, 1012 spectra of non-stellar objects were collected,
which span the redshift range $0 \le z \le 0.85$\footnote{Six quasars at $z=1.09$, 1.32, 2.01, 2.04, 2.14, and 3.16 were also detected.}, of which 104 result
to be multiple observations.
This gives an intrinsic error on spectroscopic redshifts
equal to $254 \pm 49~\mathrm{km~s^{-1}}$, in good agreement
with our previous determination and the VVDS one
(cf. Sect.~\ref{a2744specmem}).
In conclusion, the catalogue of VIMOS galaxy spectra
in the region of RXCJ\,2308.3-0211 is made of 902 robustly identified objects.
No additional spectroscopic redshifts were found in {\sl NED}.

According to the procedure of assigning cluster membership
previously described, the final catalogue of spectroscopically confirmed
members of RXCJ\,2308.3-0211 comprises 269 galaxies.
The fit of the corresponding red-sequence members gives:
\begin{displaymath}
(B-R) = (3.026\pm0.748) -(0.043\pm0.013)\times R. \pm 0.114
\end{displaymath}
Comparison of the analogous loci defined by photometric or spectroscopic
members of RXCJ\,2308.3-0211 confirms the behaviour
found for RXCJ\,0014.3-3022.
The same holds for the comparison of fits obtained for the two clusters.
Therefore, the fit of the photometric members of the red sequence
of either cluster will be used hereafter, as this is based
on larger statistics.

\begin{figure}
   \centering
   \includegraphics[width=8 cm]{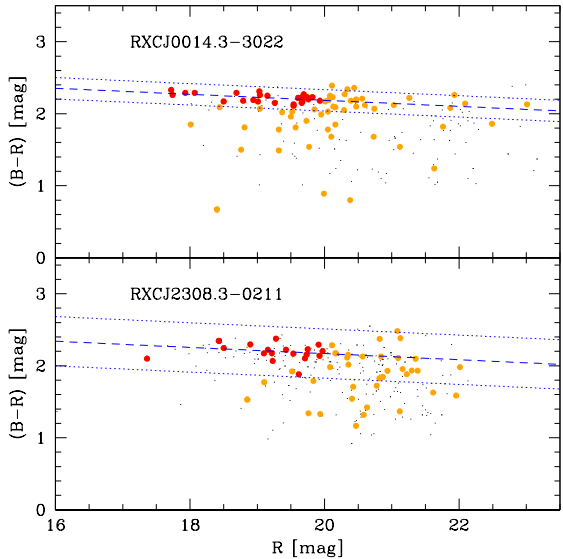}
      \caption{The same as in Fig.~\ref{rseqfield} but for the spectroscopic members of each cluster. The red sequences defined by either spectroscopic or photometric members lying within $3^{\prime}$ from the centre of each cluster are identical.}
         \label{rseqspec}
   \end{figure}

No spectrum exists for the BCG of RXCJ\,2308.3-0211,
which sits at the cluster centre (Pierini et al.~\cite{pierini08}).
Interestingly, there is a spectroscopic member with about 60\%
of the R-band luminosity of the BCG (see Fig.~\ref{rseqspec}).
It lies at a cluster-centric distance of 55$^{\prime\prime}$
(i.e. about 240 kpc at the cluster's redshift) from the cluster centre;
it was not considered in the analysis made by the previous authors.

\subsection{Cluster morphology: hot gas vs. galaxies}\label{clusmorph}

The available X-ray surface brightness distributions
(Zhang et al.~\cite{zhang06}), which witness the dynamical state of the hot gas
(Jones \& Forman~\cite{jones92}), are complemented with surface density maps
of member galaxies for the two clusters, determined as described in Appendix A.
Comparison of the two-dimensional distributions of hot gas and galaxies
improves our knowledge about substructure and large-scale structure
in and around the clusters.

The entire population of photometric member galaxies of a cluster
and its red and blue subpopulations are considered separately.
In all three cases, the same smoothing for the enhancement
of structure against a smooth background and final spike cleaning is used,
which consists in a gaussian kernel of $100^{\prime\prime}$.
Galaxy density maps were always extracted with the same resolution
(mesh width of $10^{\prime\prime}$) to directly compare results.
Spectroscopic catalogues were considered as well,
although they cover only a fraction of the region imaged with WFI.
In this case, the mesh width and smoothing scale have been increased
to 20$^{\prime\prime}$ and 250$^{\prime\prime}$, respectively,
to compensate for the poorer statistics, and, thus,
to ensure convergence and avoid occurrence of spikes produced by undersampling.
Table 4 gives results from the calculation
of the smooth background calculation for all catalogues used.
In all cases, the RMS value of the smooth background is used
as standard unit to estimate the overdensity associated with a given structure.

\begin{table}
\label{tablebg}      
\centering                          
\begin{tabular}{c c c c}        
\hline                
catalogue & N. objects & Mean BG & RMS \\    
\hline                        
   RXCJ0014.3-3022 (all) & 3698 & 32.37 & 10.31 \\      
   RXCJ0014.3-3022 (red) & 1409 & 9.70 & 4.28 \\
   RXCJ0014.3-3022 (blue) & 2289 & 21.91 & 6.90 \\
   RXCJ0014.3-3022 (spec.) & 190 & 1.90 & 3.03 \\
\hline
   RXCJ2308-0211 (all) & 4109 & 44.72 & 14.74 \\      
   RXCJ2308-0211 (red) & 2361 & 23.26 & 9.52 \\
   RXCJ2308-0211 (blue) & 1748 & 20.10 & 7.52 \\
   RXCJ2308-0211 (spec.) & 269 & 1.18 & 5.01 \\
\hline                                   
\end{tabular}
\caption{Background evaluation results for the different subcatalogues of cluster members, for both clusters. Values are always given in units of number of galaxies per square Mpc.}             
\end{table}

\subsubsection{RXCJ\,0014.3-3022}\label{a2744morph}

The complex morphology of RXCJ\,0014.3-3022 at different scales
was illustrated by B\"ohringer et al. (\cite{boehringer06}) and BPB07,
with different tracers and techniques.
In particular, there is evidence for two extended filamentary structures
of galaxies stemming out of the main body of this merging cluster,
one heading to the North-West (NW), the other to the South (S).
Here we use the same subregions identified in BPB07, i.e.: the cluster core,
the two filaments, and a neighbouring, low-density region
assumed to represent the coeval ``field'' environment unaffected by the cluster.
The statistics of the field is used to determine significant overdensities.
The geometrical criteria (in addition to those on observed magnitude
and photo-$z$) behind the definition of each subregion
(and corresponding subcatalogue) are listed in Table 5.
There the criteria are expressed as local polar coordinates with respect to 
the cluster centre, with the cluster-centric distance expressed in arcminutes,
the polar angle in degrees (with the zero angle towards E and increasing 
counterclockwise w.r.t. the observer) and the enclosed area in square arcminutes.

\begin{figure*}
   \centering
   \includegraphics[width=12truecm]{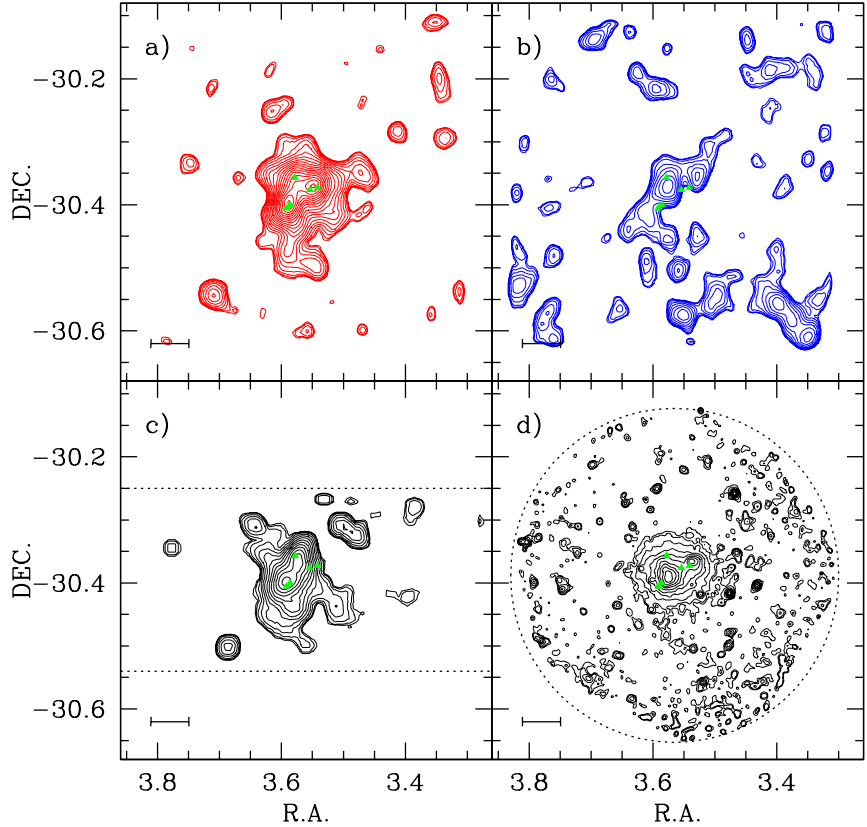}
      \caption{Galaxy density maps and X-ray image for RXCJ\,0014.3-3022. Panels a) and b) show, respectively, the density maps of red and blue photometric members. Panel c) shows the density map of spectroscopic members instead, where dotted lines mark the approximate extent of the VIMOS combined pointings. Panel d) shows the distribution of the X-ray emitting, hot gas, where a dotted circle marks the approximate FOV of XMM-\emph{Newton}. Green triangles mark the positions of the BCGs of every galaxy clumps. In all maps, reproduced contours have a significance of at least $5 \sigma$ w.r.t. the background. The lower left horizontal bar has a size of 1 Mpc at the cluster's redshift.}
         \label{dmaps0014}
   \end{figure*}

The main body of RXCJ\,0014.3-3022 is elongated in the NW-SE direction,
which is where the two main clumps of galaxies lie,
but it also exhibits lots of scattered extensions.
The most outstanding ones are the two filaments to the NW and S,
which can be traced out to about 9$^{\prime}$ (i.e., 2.44 Mpc)
and certainly extend beyond the imaged region of the cluster,
and, thus, beyond the X-ray $R_{200}$ (2.52 Mpc, Zhang et al.~\cite{zhang06}).
These two elongated structures host about the same number of galaxies,
and as many as the core, which occupies an area four times smaller though.
However, the S filament stands out more clearly as a twisted chain
of high-density lumps, whereas the NW filament appears
as a smoother distribution of slightly overdense peaks
(see Fig.~\ref{dmaps0014}).
Unsurprisingly, red galaxies highlight the main body of the cluster,
but they also trace the inner part of the filaments.
In addition to the two central clumps each associated with BGC pairs
(see Pierini et al.~\cite{pierini08}), a third, less dense clump
of red galaxies lies slighly to the N w.r.t. the cluster centre:
it hosts a fifth BCG, as anticipated (see Sect.3.2.2).
An additional overdensity of red galaxies (w.r.t. to the field)
lies to the SE; it appears as detached from the main body of the cluster.

The overall picture is confirmed by the distribution
of the spectroscopically confirmed cluster members.
Although the available spectroscopy does not extend to the outer edges
of the filaments, at least the inner part of the S filament
is traced by spectroscopic cluster members.
Remarkably the same conclusion can be reached
from the X-ray surface brightness distribution (Fig.~\ref{dmaps0014})
and entropy map (B\"ohringer et al.~\cite{boehringer06}) of the hot gas.
Unsurprisingly, filaments are better traced by blue galaxies.
Nevertheless, an overdensity of blue galaxies (w.r.t. the field)
lies exactly between the three clumps of red galaxies associated with BCGs
previously described.
Its presence is consistent with the findings of Couch et al. (\cite{couch98}).
In addition, two regions with a less marked overdensity of blue galaxies
are discovered: they just trail the two main clumps of red galaxies
in the main body of the cluster.
One of them is led by an X-ray bow-shock detected with {\sl Chandra}
(see Kempner \& David~\cite{kempner04}) and associated with
the NW subcomponent of this merging cluster.
As discussed in BPB07, this shock may shield infalling galaxies
from the surrounding ICM, which enables them to continue
their normal star formation activity.

\begin{table}
\label{regs0014}      
\centering                          
\begin{tabular}{c c c c c}        
\hline                
Region & Radius & Angle & Objects & Area \\    
\hline                        
  Cluster core & 0'$\leq$r$\leq$3' & 0$\leq\theta\leq$360 & 304 & 9$\pi$ \\      
  NW filament & 3'$<$r$\leq$15' & -10$\leq\theta\leq$50 & 359 & 36$\pi$ \\
  S filament  & 3'$<$r$\leq$15' & 260$\leq\theta\leq$320 & 374 & 36$\pi$ \\
  BG field  & 6'$<$r$\leq$18' & 120$\leq\theta\leq$210 & 488 & 72$\pi$ \\
\hline                                   
\end{tabular}
\caption{Definition of regions of interest in RXCJ0014.3-3022. Areas are expressed in square arcminutes.}
\end{table}
%

\subsubsection{RXCJ\,2308.3-0211}\label{a2537morph}

In X-rays RXCJ\,2308.3-0211 appears as an azimuthally symmetric,
centrally concentrated system, as typical of cool-core clusters.
For the region of this cluster we define five subregions (see Table 6):
the cluster core, two outer, concentric circular annuli,
and two low-density regions (i.e., ``fields''),
We note that the total circular area occupied by the cluster,
defined as the sum of the core and the two annular regions,
contains 980 galaxies within a radius of $9^{\prime}$.
For comparison, RXCJ\,0014.3-3022 hosts 1016 galaxies
within the same cluster-centric distance.
Furthermore, the total area of the field regions for RXCJ\,2308.3-0211
is equivalent to the area of the field for RXCJ\,0014.3-3022;
it also contains the same number of objects.

Consistently with the cool-core nature of the cluster,
the distribution of red galaxies is centrally concentrated,
the BCG itself sitting at the cluster centre.
However, the optical morphology of the cluster is not regular,
as confirmed by the two-dimensional distribution
of spectroscopic cluster members.
Furthermore, two pronounced overdensities of red galaxies
lie to the N of RXCJ\,2308.3-0211, about $15^{\prime}$
(i.e., $\sim 4.1~\mathrm{Mpc}$) away from it.
They are well beyond the X-ray $R_{200}$ of the cluster (2.44 Mpc; Zhang et al. 2006)
and are separated by $\sim 4^{\prime}$ (i.e., about 1.2 Mpc).
These two structures are likely groups at the same redshift
of RXCJ\,2308.3-0211; they seem to be detected in X-rays as well,
although they lie at the edge of the XMM-{\sl Newton} image.
Both the red galaxy density distribution and the two-dimensional distribution
of spectroscopic cluster members suggest the existence of a bridge
between these two groups and RXCJ\,2308.3-0211. This also compares well 
with the weak lensing analysis in Dahle et al. (\cite{dahle02}), where 
the mass distribution is seen to be elongated towards N.
An additional overdensity of spectroscopic members eastwards
of the main cluster finds correspondence in the red galaxy map
and in the X-ray image.
A Dressler--Schechtman analysis (Dressler \& Schechtman~\cite{dressler88})
suggests that this structure is likely a group which is still not connected
to the cluster (Braglia et al. 2009b, in prep.).
Finally, there are extremely few overdense structures traced by blue galaxies,
across the WFI image: the main ones form a sequence of blobs
jutting out of the cluster centre in the NE direction.

\begin{table}
\label{regs2308}      
\centering                          
\begin{tabular}{c c c c c}        
\hline                
Region & Radius & Angle & Objects & Area \\    
\hline                        
  Cluster core & 0'$\leq$r$\leq$3' & 0$\leq\theta\leq$360 & 258 & 9$\pi$ \\      
  Inner outskirts & 3'$<$r$\leq$6' & 0$\leq\theta\leq$360 & 293 & 27$\pi$ \\
  Outer outskirts  & 6'$<$r$\leq$9' & 0$\leq\theta\leq$360 & 429 & 45$\pi$ \\
  BG field (1)  & 9'$<$r$\leq$19' & 200$\leq\theta\leq$240 & 255 & 31.1$\pi$ \\
  BG field (2)  & 9'$<$r$\leq$19' & 300$\leq\theta\leq$350 & 228 & 38.9$\pi$ \\
\hline                                   
\end{tabular}
\caption{Definition of regions of interest in RXCJ2308.3-0211. Areas are expressed in square arcminutes.}
\end{table}
%

    \begin{figure*}
   \centering
   \includegraphics[width=12truecm]{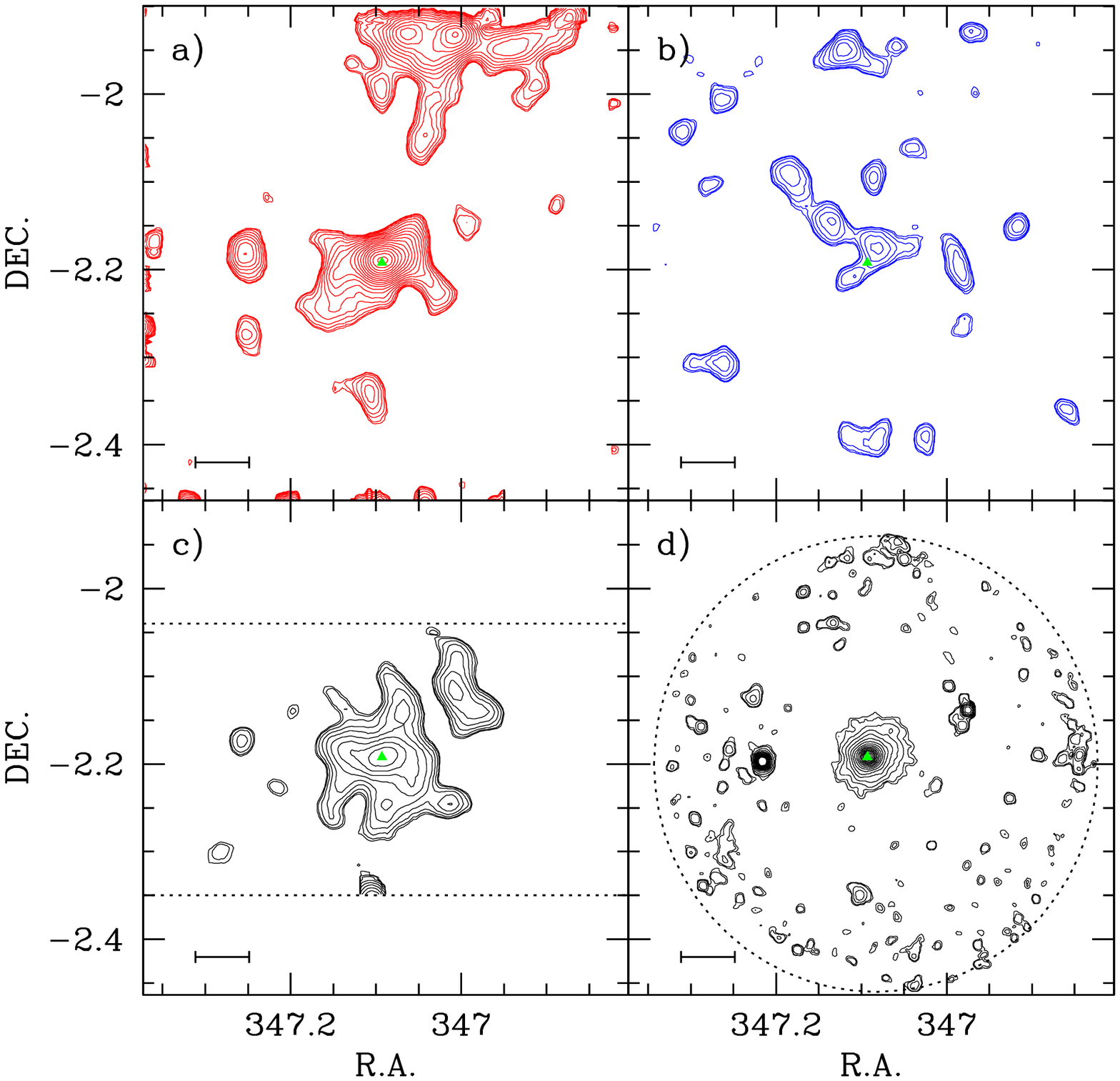}
      \caption{The same as in Fig.~\ref{dmaps0014} but for RXCJ\,2308.3-0211.}
         \label{dmaps2308}
   \end{figure*}

\subsection{Spectral indices: global behaviour of galaxy populations}\label{indices}

The strength of the 4000\,\AA~break ($D4000$) and the equivalent widths
of the $\mathrm{H_\delta}$ photospheric absorption line
and the $[OII]$ nebular emission line ($W_0 \mathrm{(H_\delta)}$
and $W_0 \mathrm{([OII])}$, respectively) could be robustly determined
from VIMOS spectra with $S/N > 4$ for galaxies at $z \sim 0.3$.
An analysis based on these spectral features can constrain
the star-formation histories of individual cluster members
(e.g. Balogh et al.~\cite{balogh99}; throughout this paper we will use the same definition 
adopted in there,where $W_0\mathrm{([OII])}$ is positive when in emission and 
$W_0 \mathrm{(H_\delta)}$ is positive in absorption).
In fact, the spectral index $D4000$ is particularly sensitive
to the presence of old (i.e., $\ge 1$--$2~\mathrm{Gyr}$), passively evolving
stellar populations.
Conversely, $W_0 \mathrm{(H_\delta)}$ is sensitive to episodes
of star formation in the last Gyr, since it measures the presence
of A-type stars, whereas $W_0 \mathrm{([OII])}$ estimates
the present star formation activity (i.e., the SFR on timescales
of $\le 0.01~\mathrm{Gyr}$).
The combined use of the diagnostic diagrams
$W_0 \mathrm{(H_\delta)}$--$W_0 \mathrm{([OII])}$
and $W_0 \mathrm{(H_\delta)}$--$D4000$ limits misinterpretations
arising from non introducing corrections for intrinsic extinction by dust
or additional line emission affecting $W_0 \mathrm{([OII])}$
and $W_0 \mathrm{(H_\delta)}$, respectively.

\begin{figure}
   \centering
   \includegraphics[width=8 cm]{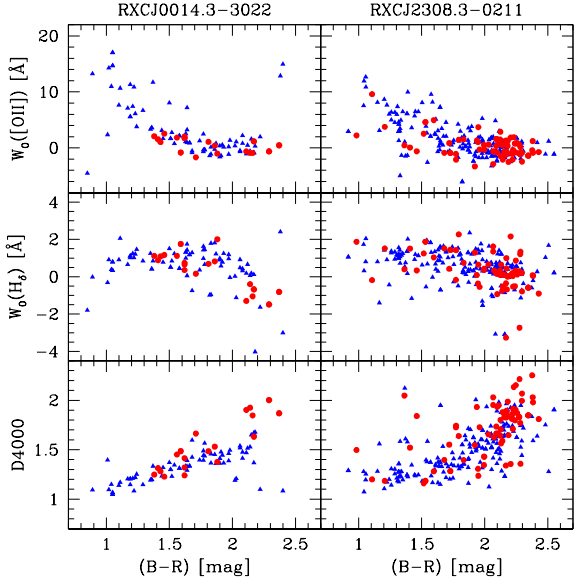}
      \caption{Spectral indices $W_0 \mathrm{(H_\delta)}$, $W_0 \mathrm{([OII])}$, and $D4000$ (see text) as a function of $(B-R)$ colour (observed frame) for spectroscopic members of RXCJ\,0014.3-3022 (left column) and of RXCJ\,2308.3-0211 (right column). Triangles and circles mark sub-L$^{\star}$ (i.e., with $L_\mathrm{R} < L_\mathrm{R}^{\star}$) and luminous galaxies, respectively.}
         \label{indexcolours}
   \end{figure}

The behaviour of each spectral index as a function
of the observed $(B-R)$ colour (uncorrected for internal extinction)
for the spectroscopic members of each cluster is reproduced
in Fig.~\ref{indexcolours}.
Unsurprisingly, galaxies with redder $(B-R)$ colours tend to exhibit
lower values of $W_0 \mathrm{([OII])}$ (i.e., less emission in the [OII] line)
and larger values of $D4000$.
Hence, they have experienced less star-formation activity
since the last Gyr or so, and, thus, tend to be dominated
by old, passively evolving stellar populations.
Interestingly, this holds irrespectively of the galaxy luminosities.

In analogy with previous spectral index analyses
(e.g. Barger et al.~\cite{barger96}; Balogh et al.~\cite{balogh99}),
we empirically determine threshold values for the three spectral indices
in order to discriminate different star-formation regimes.
In particular, red sequence galaxies identified photometrically
(cf. Sect.~\ref{catphot} and \ref{catspec}) are used to determine
mean and RMS of $W_0 \mathrm{([OII])}$, $W_0 \mathrm{(H_\delta)}$,
and $D4000$ that are assumed to represent galaxies dominated
by old, passively evolving stellar populations.
These are defined as the objects falling within $3 \sigma$ from the mean
of $W_0 \mathrm{([OII])}$ or $D4000$, which operationally translates into:
$W_0 \mathrm{([OII])} < 4.85$\,\AA~in the
$W_0 \mathrm{([OII])}$--$W_0 \mathrm{(H_\delta)}$ diagnostic diagram,
or $D4000 \geq 1.4$ in the $D4000$--$W_0 \mathrm{(H_\delta)}$ one.
These thresholds are well consistent with those adopted
by Balogh et al. (\cite{balogh99}) or Dressler et al. (\cite{dressler99}),
according to their samples and data (i.e., $W_0 \mathrm{([OII])} = 5$\,\AA~and
$D4000 = 1.45$, respectively).
Following these authors, we adopt $W_0(H_{\delta}) = 0$\,\AA~as a threshold
between normal star-forming and (short) starburst galaxies.

In absence of detected objects with extreme values
of $W_0 \mathrm{(H_{\delta})}$ in our samples,
the spectroscopically confirmed member galaxies of our clusters
belong to one of the following three classes (see also Tables~\ref{defo2hd}
and \ref{defd4hd}; cf. Balogh et al. (\cite{balogh99})).

{\it Passive galaxies (PEV)}: systems currently not undergoing
traceable star formation activity, mainly with E or S0 spectral type
(and morphology).
They are defined by $W_0 \mathrm{(H_{\delta})} \leq 5$\,\AA~and
$W_0 \mathrm{([OII])} < 4.85$\,\AA, or $D4000 \geq 1.4$.

{\it Star-forming galaxies (SF)}: systems undergoing
significant star formation activity since at least several hundred
million years, mainly identified as late-type galaxies
(spirals and irregulars).
They are identified by $0 \leq W_0 \mathrm{(H_{\delta})} \leq 5$\,\AA~and
$W_0 \mathrm{([OII])} \geq 4.85$\,\AA, or alternatively $D4000 < 1.4$.

{\it Short starburst galaxies (SSB)}: systems currently undergoing
a short-lived, intense episode of star formation (i.e., $\leq$200 Myr),
where nebular emission is strong.
They have $W_0 \mathrm{(H_{\delta})} < 0$\,\AA~and
$W_0 \mathrm{([OII])} \geq 4.85$\,\AA, or alternatively $D4000 < 1.4$.

Although none of the cluster members fall into the following definition,
we want to define here an additional class which will be useful in our analysis,
the so-called {\it post-starburst galaxies}
(Dressler and Gunn \cite{dressler83}; Poggianti et al. \cite{poggianti99}).
These systems show negligible emission lines but strong Balmer absorption; this
suggests that star formation activity in these systems ended abruptly in the recent past.
According to Poggianti et al. (1999) and our previous definitions,
we define these systems as having $W_0 \mathrm{(H_{\delta})} \geq 5$\,\AA
and $W_0 \mathrm{([OII])} < 4.85$\,\AA.

\begin{figure}
   \centering
   \includegraphics[width=8 cm]{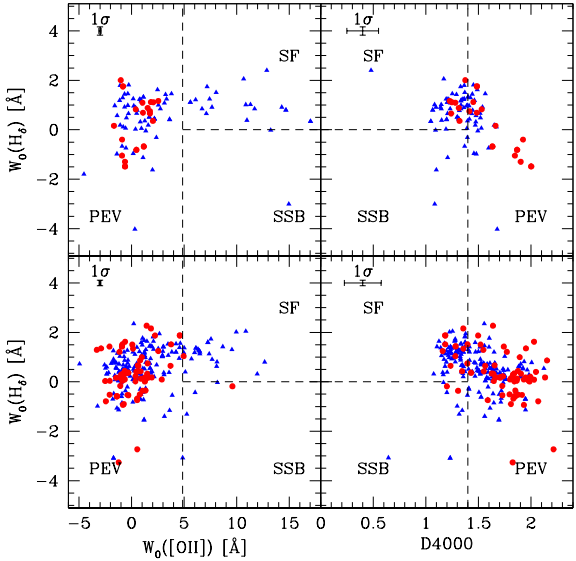}
      \caption{Spectral index diagnostic diagrams for spectroscopic members of RXCJ\,0014.3-3022 (top row) and of RXCJ\,2308.3-0211 (bottom row). In each panel, $1 \sigma$ uncertainties are shown, whereas dashed lines mark thresholds for discriminating different star-formation regimes (see Tables \ref{defo2hd} and \ref{defd4hd}). Symbols are the same as in Fig.~\ref{indexcolours}.}
         \label{indexall}
   \end{figure}

The diagnostic diagrams $W_0 \mathrm{([OII])}$--$W_0 \mathrm{(H_\delta)}$
and $D4000$--$W_0 \mathrm{(H_\delta)}$ establish that RXCJ\,0014.3-3022
and RXCJ\,2308.3-0211 host overall different galaxy populations
(see Fig.~\ref{indexall}).
In fact, a diagnostic two-dimensional Kolmogorov-Smirnov test
(Fasano \& Franceschini~\cite{fasano87}; Peacock~\cite{peacock83}),
executed for the distribution of members of either cluster
in the two previous diagnostic diagrams, concludes that
the probability that the two distributions are drawn
from the same parent population is absolutely negligible:
0.02\% for the $D4000$--$W_0 \mathrm{(H_\delta)}$ plane,
and 0.41\% for the $W_0 \mathrm{([OII])}$--$W_0 \mathrm{(H_\delta)}$ plane.
The very significant difference between the galaxy populations
of the two clusters is not due to selection effects,
since the selection function of spectroscopic targets was the same
(cf. Sect.~\ref{dataspec}).

\begin{figure}
   \centering
   \includegraphics[width=8 cm]{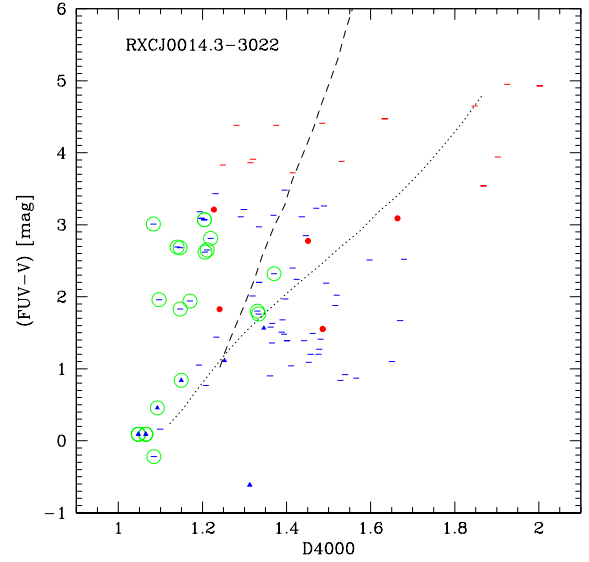}
   \includegraphics[width=8 cm]{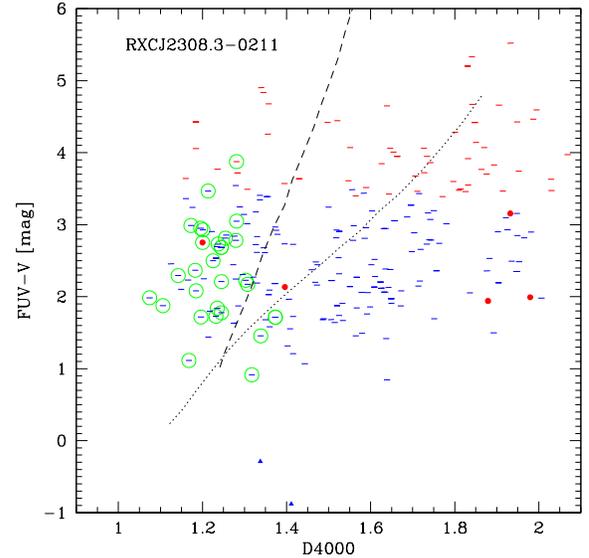}
      \caption{Rest-frame $(FUV-V)$ colour as a function of $D4000$ for the spectroscopic members of RXCJ\,0014.3-3022 (top) and of RXCJ\,2308.3-0211 (bottom). In each panel, symbols are the same as in Fig.~\ref{indexcolours}. In addition, red and blue bars mark lower limits on $(FUV-V)$ for luminous and sub-L$^{\star}$ galaxies, respectively. Furthermore, green circles mark galaxies with significant [O\,II] emission (i.e., $W_0 (\mathrm{[O\,II]}) \geq 4.85$\,\AA). Finally, dashed and dotted lines mark, respectively, models corresponding to a starvation or a truncation evolutionary track, similarly to Moran et al. (\cite{moran07}).}
         \label{d4uvplots}
   \end{figure}

This is confirmed by the different distribution
of spectroscopic cluster members in the $D4000$--$(FUV-V)$ colour (rest frame\footnote{GALEX FUV and WFI V-band magnitudes of individual objects are $k$-corrected according to the best-fit templates in \emph{HyperZ}.}) plane (cf. Moran et al.~\cite{moran07}), as shown in Fig. \ref{d4uvplots}.
There galaxies are flagged as passive or star-forming according
to their values of $W_0 \mathrm{([OII])}$.
Out of 101 (269) spectroscopic members of RXCJ\,0014.3-3022
(RXCJ\,2308.3-0211), 12 (7) were detected by GALEX in the FUV\footnote{Most of the FUV-detected galaxies in RXCJ\,0014.3-3022 lie close to the cluster core and in the region of the S filament, in correspondence of the clumps of blue galaxies discussed in Sect.~\ref{clusmorph}.}.
It is evident that not only the fraction of FUV-detected member galaxies,
but also the overall distribution of member galaxies with lower limits
on $(FUV-V)$ is different.
In particular, the bulk of spectroscopic members of RXCJ\,2308.3-0211
does not exhibit $(FUV-V) < 2$ (rest frame), whatever the value of $D4000$,
whereas RXCJ\,0014.3-3022 hosts a non-negligible population of galaxies
with $D4000<1.6$~and $(FUV-V) < 2$.
We attribute the strong differences in galaxy populations
between RXCJ\,0014.3-3022 and RXCJ\,2308.3-0211 to the largely different
dynamical states and morphologies of the two clusters
(see Sect. \ref{clusmorph} and Braglia et al. 2009b, in prep.).

Finally, we note that both clusters host a small but odd population
of galaxies with $(B-R) \geq 2$ and either $D4000 < 1.4$
or $W_0 \mathrm{([OII])} \geq 4.85$\,\AA.
In RXCJ\,0014.3-3022 these few objects are sub-L$^{\star}$,
whereas there are some luminous ones in RXCJ\,2308.3-0211.
We tentatively interpret this odd class of red galaxies with evidence
of recent or present star-formation activity as dusty galaxies
with very different SFRs (cf. Verdugo et al.~\cite{verdugo08}).
They are not post-starburst galaxies though,
since they do not show a large absorption in the $\mathrm{(H_\delta)}$ line (i.e. $\geq 5$).
There is now substantial observational evidence from IR observations
that cluster galaxies can experience strongly obscured star-formation activity
(e.g. Metcalfe et al.~\cite{metcalfe05}; Saintonge et al.~\cite{saintonge08}, and references therein).
This population of dust-obscured star-forming galaxies seems particularly abundant in galaxy clusters
undergoing merging and strong episodes of mass accretion (e.g. Coia et al.~\cite{coia05}; Tran et 
al.~\cite{tran05}; Loh et al.~\cite{loh08}), with little dependence on the mass of the parent system
as they are also observed in groups (Wilman et al.~\cite{wilman08}). A large fraction of these dusty galaxies
are classified as early-type spirals (e.g. Biviano et al.~\cite{biviano04}; Wilman et al.~\cite{wilman08}),
in agreement with the average spectral type of our red, star-forming population.

\begin{table}
\caption{Definitions in the [$W_0(OII)$, $W_0(H_{\delta})$] plane}             
\label{defo2hd}      
\centering                          
\begin{tabular}{c c c}        
\hline                 
Definition & $W_0(OII)$ & $W_0(H_{\delta})$ \\    
\hline                        
PEV (passive) & $<4.85$ & $\leq5$ \\      
SF (star-forming) & $\geq4.85$ & $\in$[0, 5] \\
SSB (short starburst) & $\geq4.85$ & $<0$ \\
\hline                                   
\end{tabular}
\end{table}
%

%
\begin{table}
\caption{Definitions in the [$D4000$, $W_0(H_{\delta})$] plane}             
\label{defd4hd}      
\centering                          
\begin{tabular}{c c c}        
\hline                 
Definition & $D4000$ & $W_0(H_{\delta})$ \\    
\hline                        
PEV (passive) & $\geq1.4$ & $\leq5$ \\      
SF (star-forming) & $<1.4$ & $\in$[0, 5] \\
SSB (short starburst) & $<1.4$ & $<0$ \\
\hline                                   
\end{tabular}
\end{table}
%

%

%

\subsection{Spectral indices: local behaviour of galaxy populations}

\begin{figure*}
   \centering
   \includegraphics[width=12truecm]{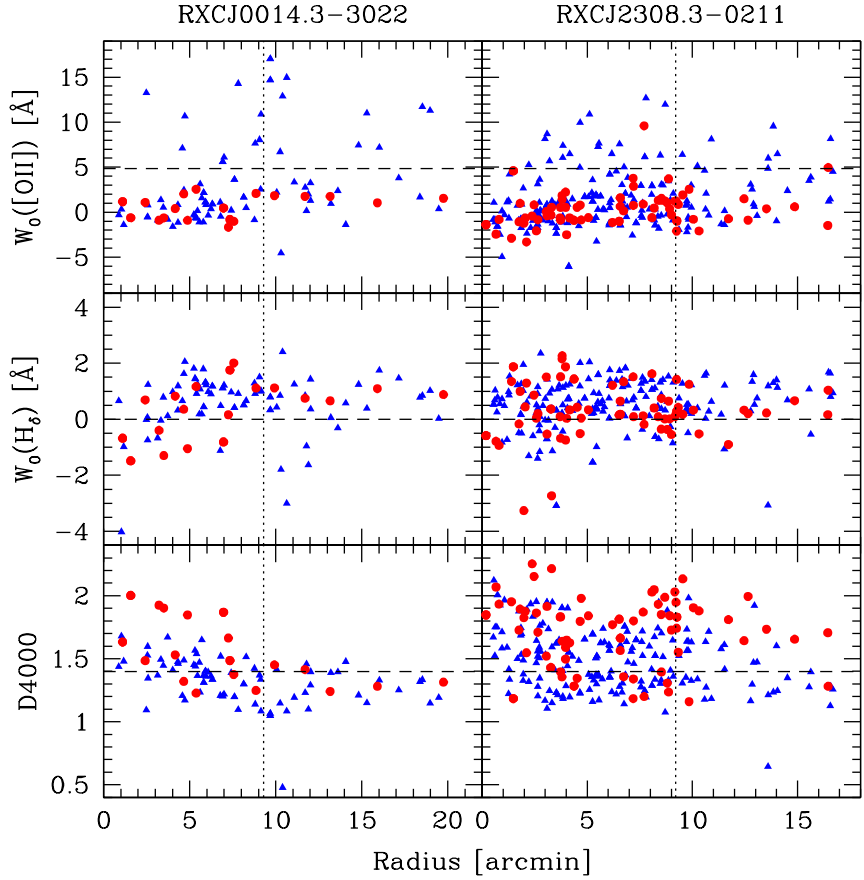}
      \caption{Behaviour of spectral indices as a function of cluster-centric distance for spectroscopic members of RXCJ\,0014.3-3022 (left column) and of RXCJ\,2308.3-0211 (right column). Symbols are the same as in Fig.~\ref{indexcolours}. In each panel, the horizontal dashed line marks the characteristic threshold for the given index, whereas the vertical dotted line marks the distance corresponding to $R_{200}$.}
         \label{ewrad}
   \end{figure*}

Here we investigate if differences in galaxy populations
between RXCJ\,0014.3-3022 and RXCJ\,2308.3-0211
arise in well defined regions of the two clusters.
Firstly, we analyse the behaviour of the spectral indices
of spectroscopically confirmed members
as a function of the cluster-centric distance (see Fig.~\ref{ewrad}).
In RXCJ\,0014.3-3022, spectroscopic members tend to exhibit
$D4000 < 1.4$~when their cluster-centric distances approach
or exceed the value of $R_{200}$.
The increase of the fraction of star-forming systems
towards and beyond $R_{200}$ is confirmed by the corresponding behaviours
of $W_0 \mathrm{([OII])}$ and $W_0 \mathrm{(H_\delta)}$.
This is consistent with the existence of a sharp peak in the fraction
of blue-to-red galaxies along the two filaments (see BPB07)
especially the southern one, which is poorly sampled
by the available VIMOS data (see Fig.~\ref{vimosmasks}).
Conversely, the combined behaviour of the same spectral indices
for spectroscopic members of RXCJ\,2308.3-0211
as a function of cluster-centric distance shows that
passive galaxies dominate everywhere in this cluster.
Furthermore, luminous galaxies
(i.e., with $L_\mathrm{R} \geq L_\mathrm{R}^{\star}$
tend to be passive already beyond $R_{200}$ in RXCJ\,2308.3-0211,
whereas passive, luminous galaxies almost exclusively populate
regions within $R_{200}$ in RXCJ\,0014.3-3022.
Analogous differences exist when considering the population
of sub-L$^{\star}$ galaxies in the two clusters.
This suggests that galaxies already tend to be passive
and, thus, populate the red sequence only outside the virialized region
in RXCJ\,2308-0211, in agreement with previous findings
for other clusters (Kodama et al.~\cite{kodama01},~\cite{kodama03};
Balogh et al.~\cite{balogh97}).
In particular, Balogh et al. (\cite{balogh97}) found that
star formation in cluster galaxies is generally suppressed
(w.r.t. the field) out to $2R_{200}$.
Conversely, galaxies tend to mostly populate the red sequence
inside the virialized region in RXCJ\,0014.3-3022.

\begin{figure*}
   \centering
   \includegraphics[width=12truecm]{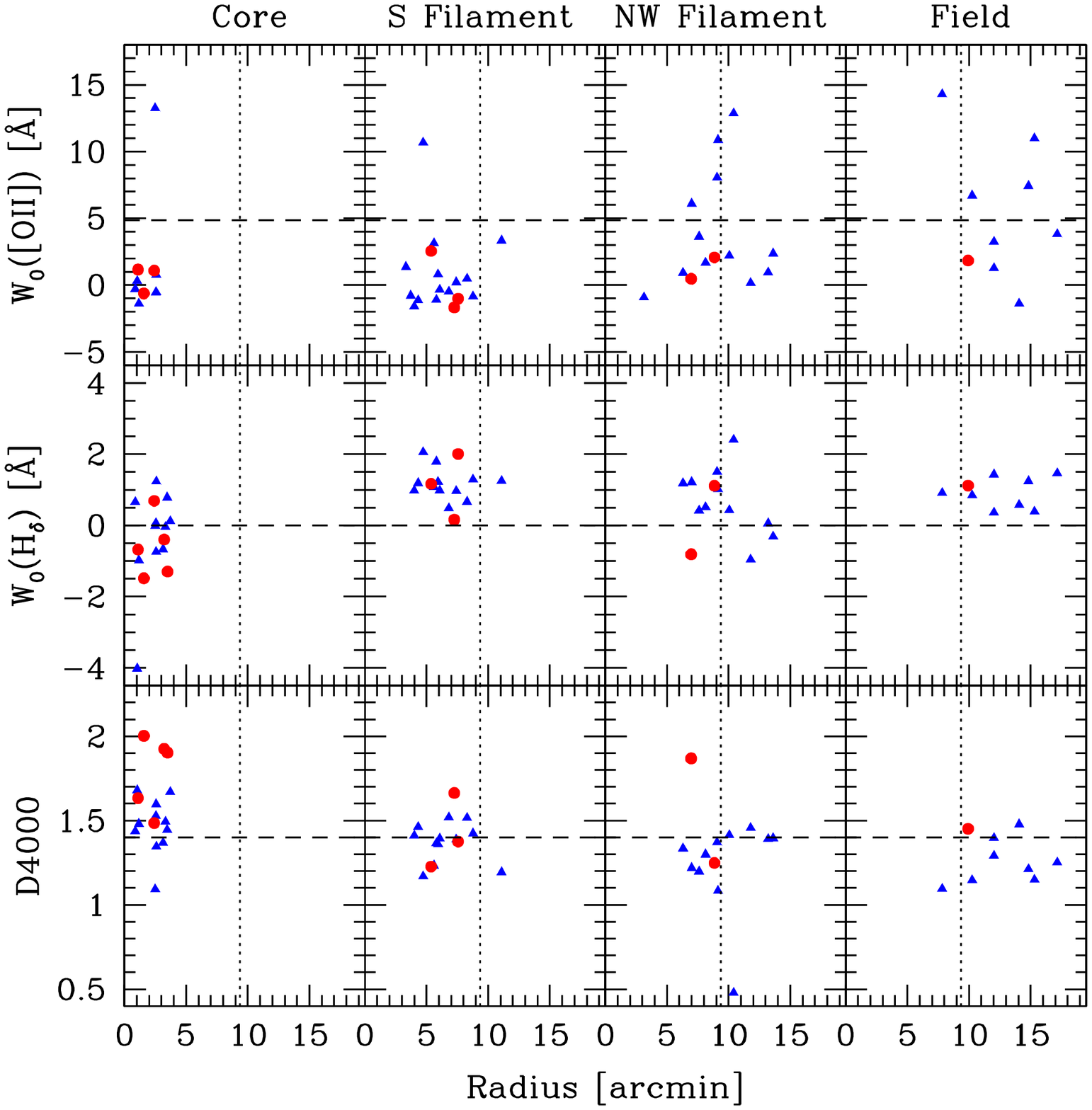}
      \caption{Behaviour of spectral indices as a function of cluster-centric distance for different regions of RXCJ\,0014.3-3022 (see Table 5). Symbols and lines are the same as in Fig.~\ref{ewrad}.}
         \label{ewrad0014}
   \end{figure*}

To better understand these differences, we investigate subregions
of the two clusters as well as their neighbouring fields.
As for RXCJ\,0014.3-3022 (Fig.~\ref{ewrad0014}), spectroscopic members
in the core region are mostly passive at variance with the galaxy population
in the neighbouring field.
This contains a mixture of passive and mildly-to-highly star-forming systems.
Nevertheless, there is additional evidence of the presence
of (mostly faint) galaxies with intense star-formation activity
in the cluster core (see Couch et al.~\cite{couch98}).
Spectroscopic members associated with the two filaments of RXCJ\,0014.3-3022
exhibit properties in between those of the galaxy populations
in the cluster core and the neighbouring field.
Unfortunately, the present VIMOS data do not extend much beyond $R_{200}$ in the direction of the S filament, where most of the ``flaming giants'' discovered by BPB07 lie.

As for RXCJ\,2308.3-0211, the star-formation activity
of spectroscopic members appears as low everywhere, but tend to increase
towards $R_{200}$ (cf. Fig. \ref{ewrad}, right panel).
Consistently, $D4000$ tends to decrease from the core
to the outskirsts of the cluster.
Star-formation activity appears to be almost exclusively present
in sub-L$^{\star}$ galaxies.
This picture is in agreement with previous results in the literature
(e.g. Balogh et al.~\cite{balogh99}).
However, the paucity of star-forming galaxies,
not only among the luminous ones, is evident also in the neighbouring field.

%

\section{Discussion}\label{discussion}

RXCJ\,0014.3-3022 and RXCJ\,2308.3-0211 are two clusters at $z \sim 0.3$
with comparable X-ray luminosities 
\footnote{respectively 1.29 and 1.02 $\times 10^{45}$~ergs~s$^{-1}$} (thus, likely, total masses),
but with opposite X-ray morphologies (i.e., dynamical states):
the former is a merging system, the latter a regular,
centrally concentrated one.
The optical morphologies of both clusters are more complex than the X-ray ones.
In addition to the two X-ray subcomponents, RXCJ\,0014.3-3022 exhibits
two extended filaments (one of which is tentatively detected in X-rays)
and a core rich in substructure, as traced by galaxies with photometric
or spectroscopic redshift consistent with that of the cluster.
Furthermore, its core hosts two visual pairs of BCGs,
each associated with a displaced X-ray subcomponent
(see Pierini et al.~\cite{pierini08} for a discussion),
and an additional equally luminous BCG sitting in a third overdense region.
On the other hand, the X-ray centroid of RXCJ\,2308.3-0211 coincides
with the position of its BCG, as expected for a cool-core cluster,
but photometric or spectroscopic cluster members shape a core region
which is less centrally symmetric than in X-rays.
The difference between optical and X-ray shape can be well explained
in terms of the different properties of the various cluster components.
The gas particles have isotropic motions, thus follow isopotential contours which are
smoother than the distribution of galaxies and dark matter, which instead show
non-isotropic velocity distributions. Moreover, the X-ray surface brightness follows
the projected square density of the gas and thus emphasises the more regular
central region, mainly missing the irregular outskirts where the gas density is low.

In spite of the evident difference in assembly state at $z \sim 0.3$,
these clusters exhibit the same locus of galaxies
dominated by old, passively evolving stellar populations\footnote{This is likely even though no correction for dust attenuation is applied to the photometry of star-forming galaxies.}.
The red sequence of each cluster is well populated, whether it is traced
by photometric or spectroscopic members.

However, the star-formation history of the average galaxy is quite different
for the two clusters, as confirmed by their spectroscopic members.
Combining information from the distributions of these galaxies
in the $W_0 \mathrm{([OII])}$--$W_0 \mathrm{(H_\delta)}$,
$D4000$--$W_0 \mathrm{(H_\delta)}$, and $D4000$--$(FUV-V)$ planes\footnote{This minimises the impact of dimming and reddening effects produced by dust in star-forming galaxies.} offers the consistent picture that star-formation activity
is rapidly moving towards a quiescent state, if it is not already absent,
across the virialized region of RXCJ\,2308.3-0211,
whatever the R-band luminosity of its spectroscopic members.
This is confirmed by comparison of the distributions
in the $D4000$--$(FUV-V)$ plane of real galaxies
and models corresponding to evolutionary scenarios for either starvation
or (abrupt) truncation of the star-formation activity
(Moran et al.~\cite{moran07}).
In both clusters, we recover the standard scenario that
the overall stellar population in galaxies tends to become younger
towards the outskirts (e.g. Balogh et al.~\cite{balogh99}).
In RXCJ\,2308.3-0211, this is mostly due to (residual) star-formation activity
in sub-L$^{\star}$ galaxies, since even the neighbouring field contains
galaxies with $L_\mathrm{R} \geq L_\mathrm{R}^{\star}$
as red as luminous red-sequence galaxies in the cluster core.

Conversely, RXCJ\,0014.3-3022 contains a sizeable fraction
of actively star-forming systems even within $R_{200}$ and beyond.
This happens in different subregions of this merging cluster
for different reasons.
sub-L$^{\star}$ starburst systems in the cluster core
were discovered by Couch et al. (\cite{couch98});
they experience strong tidal interactions.
Both sub-L$^{\star}$ and luminous star-forming systems populate the two filaments
stemming out of the cluster main body and reaching $R_{200}$
and beyond (BPB07).
The former can be interpreted as infalling low-mass systems
that manage to keep part of their still conspicuous H\,I gas reservoir,
consistently with the generally accepted ``downsizing'' scenario
(Lilly et al.~\cite{lilly96}; Gavazzi \& Scodeggio~\cite{gavazzi96a};
Gavazzi et al.~\cite{gavazzi96b}; Cowie et al.~\cite{cowie99}),
in spite of the growing influence of the cluster environment.
The latter were interpreted by BPB07 as relatively high-mass galaxies
with residual H\,I gas in their discs experiencing ``harassment''
(cf. Moore et al.~\cite{moore96},~\cite{moore98},~\cite{moore99}).
After an average increase of star-formation activity across $R_{200}$
(BPB07), star-forming galaxies of all masses along the filaments
appear to move towards the red sequence as their cluster-centric distances
become smaller and smaller with respect to $R_{200}$.
This is consistent both with the starvation scenario
(Balogh et al.~\cite{balogh99}) and with galaxy harassment,
as well as with the later removal of residual H\,I gas
through ram pressure stripping by the ICM (e.g., Gunn \& Gott~\cite{gunn72};
Quilis et al.~\cite{quilis00}; Tonnesen \& Bryan~\cite{tonnesen08})
in the cluster core.

In conclusion, RXCJ\,2308.3-0211 appears as a massive system
which is already assembled at $z \sim 0.3$,
and sits in a large-scale environment where most galaxies
are dominated by old, passively evolving stellar populations
(cf. the two groups well beyond the cluster virial radius)
as well in its virialized region.
Conversely, the massive, merging system RXCJ\,0014.3-3022
sits in a large-scale environment where galaxies of different masses
can still form stars, as it happens in its core and filaments.
Nevertheless, galaxies dominated by old, passively evolving stellar populations
are the most frequent ones in its core, and define the same locus
as those in the core of RXCJ\,2308.3-0211, although with half the scatter.
This is consistent with the previous picture
that a larger and more heterogeneous fraction of the overall galaxy population
has stopped forming stars since long or is rapidly moving
towards a quiescent star-formation activity in RXCJ\,2308.3-0211,
whereas in RXCJ\,0014.3-3022 only the massive galaxies
populating the cluster core have done so.

This is also reflected in the different scatter observed 
for the two clusters' red sequences. 
The tighter red sequence observed in RXCJ\,0014.3-3022 can be ascribed 
to this cluster only having its pristine red sequence, i.e. its primordial 
population of old, evolved galaxies, the large fraction of observed active objects 
having not yet moved to the red sequence. 
Conversely, RXCJ\,2308.3-0211 has much less star-forming systems, 
hence the largest part of the galaxies has already moved to the red sequence. 
If, consistently with the already described picture, 
at least part of these galaxies have just recently moved to the red sequence, 
then these younger additions will increase the observed scatter 
before finally settling on the main red sequence.

Our results support an inside--out scenario
for the build-up of the red sequence (e.g. Lidman et al.~\cite{lidman08}).
They also suggest the existence of a link between assembly history
and average star-formation activity in member galaxies of clusters
with similar masses in general.
This could explain the large variation in the evolutionary phases of galaxies
in groups with similar masses since $z \sim 1$ (Tanaka et al.~\cite{tanaka08}).
Studies based on larger statistics and multiwavelength information
than in the present one are necessary to elucidate this matter.

\section{Conclusions}\label{conclusions}

The link between dynamical state and star-formation history in member galaxies
is investigated for two X-ray luminous clusters at $z \sim 0.3$, i.e.:
the merging system RXCJ\,0014.3-3022 and the centrally symmetric system
RXCJ\,2308.3-0211.
The richness of spectroscopic information and the quality of photometric data
allow differences in galaxy populations to be traced not only
as a function of cluster-centric distance (as done in most previous studies
on clusters at similar redshifts), but also of cluster substructure.
This is more evident in maps of galaxy overdensity
(w.r.t. a coeval, neighbouring ``field'')
based on the (observed frame) optical colour $(B-R)$
and photometric redshifts.

The (observed frame) $(B-R)$--$R$ colour--magnitude diagrams
of the inner regions of both clusters exhibit the same, well-defined locus
of galaxies dominated by old, passively evolving stellar populations
(the so-called red sequence).
This locus is more populated, and with a larger scatter around it,
in the regular cluster RXCJ\,2308.3-0211.

A combined analysis based on spectroscopic indices
(i.e., the equivalent widths of the [O\,II] and $\mathrm{H_\delta}$ lines
and the amplitude of the 4000\,\AA~break $D4000$)
and (rest frame) $FUV-V$ colours establishes the existence
of substantial differences between the two clusters.
In agreement with analogous studies in the literature,
the capability of forming stars increases
the larger the cluster-centric distance of a member galaxy.
There is also evidence for obscured star-formation activity
in a small fraction of member galaxies.
However, in RXCJ\,2308.3-0211, the bulk of the luminous
(i.e., with $L_\mathrm{R} \geq L_\mathrm{R}^{\star}$)
galaxy population has stopped forming stars since long,
or is evolving towards a quiescent star-formation activity:
it's mostly the sub-L$^{\star}$ galaxies which carry on forming stars
at substantial rates.
Conversely, RXCJ\,0014.3-3022 hosts luminous galaxies
with enhanced star-formation activity (w.r.t. the ``field'')
along two filamentary structures stemming out of its main body
and reaching its $R_{200}$.
Furthermore, an increased fraction of (mostly sub-L$^{\star}$)
star-forming galaxies is found along the two filaments across $R_{200}$.
Both phenomena are likely due to ``galaxy harassment''.
Finally, it was already known that sub-L$^{\star}$ starburst galaxies
populate the core of RXCJ\,0014.3-3022: they are tidally disturbed systems.

Differences in galaxy populations extend from the virialized regions
to the large-scale environments of the two clusters.
In fact, RXCJ\,2308.3-0211 sits in a region of the universe
where passive evolution of stellar populations
or quiescent star-formation activity characterises at least luminous galaxies,
as witnessed by the field and two groups at the same (photometric) redshift.
Conversely, the two extended filaments of RXCJ\,0014.3-3022
are nested inside a still ``active'' environment.
This suggests the existence of a link between assembly history
and average star-formation activity in member galaxies
for (at least) X-ray selected, massive clusters at $z \sim 0.3$.

\begin{acknowledgements}
Based on observations made with ESO VLT at the Paranal Observatory
under programme ID 169.A-0595.
Based on observations made with the ESO/MPG 2.2m telescope
at the La Silla Observatory inside the Max Planck Gesellschaft (MPG) time.
This research has made use of the NASA/IPAC Extragalactic Database (NED)
which is operated by the Jet Propulsion Laboratory, California Institute
of Technology, under contract with the National Aeronautics
and Space Administration.
F.B. acknowledges support by the International Max-Planck Research School
(IMPRS) on Astrophysics.
D.P. acknowledges support by the German \emph{Deut\-sches Zen\-trum
f\"ur Luft- und Raum\-fahrt, DLR\/} project number 50~OR~0405.
H.B. acknowledges support by \emph{The Cluster of Excellence
``Origin and Structure of the Universe''\/},
funded by the Excellence Initiative of the Federal Government of Germany,
\emph{EXC\/} project number 153.
\end{acknowledgements}

\begin{appendix}
\label{dmapalg}
\section{Density map calculation}

Starting from the main catalogue of cluster photo-z members and the two sub-catalogues of red and blue objects, a fixed-width mesh is applied to the full catalogue. The mesh width is found to partly affect the final density map, mainly by smoothing small-scale density peaks; its best value is found to be between 7 and 10 arcsec, corresponding to a physical scale of 30 to 45 kpc at the cluster's redshift. To ensure a better sampling of the cluster population, a scale of 10 arcseconds was chosen. This is a small enough scale to detect and suitably map substructures in dense regions (like the cluster core where three main prominent blobs are seen in the galaxy distribution), while ensuring a good and stable sampling of low-density regions without being heavily affected by small-scale background fluctuations. A fixed mesh width is found to better map both high-density peaks and background, while adaptive meshes proved to be too sensitive to small-scale fluctuations, with the risk of a wrong evaluation of background.

   \begin{figure}
   \centering
   \includegraphics[width=8 cm]{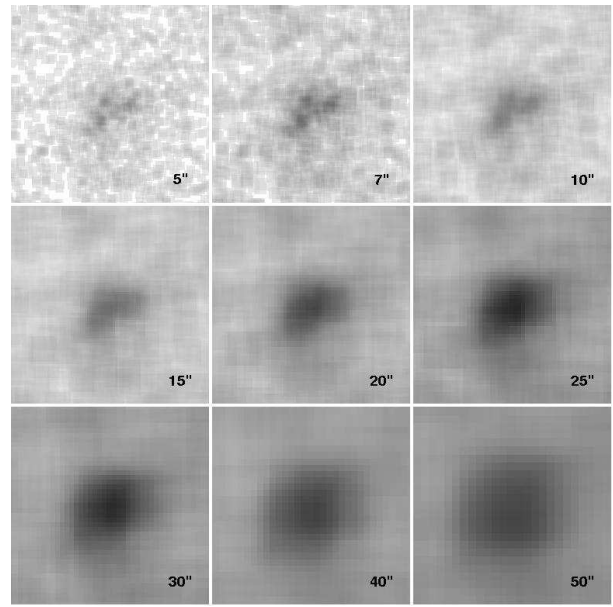}
      \caption{Example of density map for the central part of A2744. From top left to bottom right, density maps for different mesh widths are shown. The mesh width value in arcseconds is also shown for each map.}
         \label{wmesh}
   \end{figure}

Background is calculated in each mesh point by counting all objects in a fixed radius around the point. The chosen radius partly acts as a smoothing filter: this ensues a good sampling of the large-scale shape of background (so to be able to correctly evaluate background mean and RMS).
The background mean value and RMS are then calculated by fitting the background values distribution with a gaussian, assuming that the field should represent a population of randomly scattered objects with normal fluctuations. The gaussian is fitted through recursive 3-sigma clipping until convergence, which is always reached with 3 to 10 cycles. The final gaussian parameters are taken as mean and RMS of background.

   \begin{figure}
   \centering
   \includegraphics[width=8 cm]{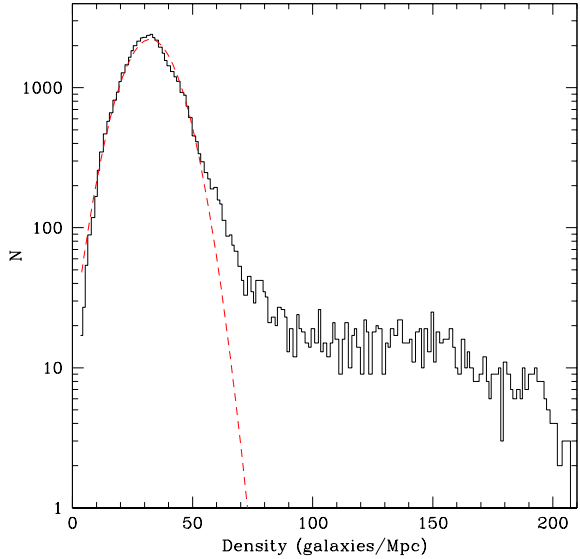}
      \caption{Background fit for a smoothing radius of 110" in the field of RXCJ\,0014.3-3022. The histogram is the count distribution in the field, with the high density tail due to overdense structures (i.e. the cluster and its filaments); the dashed line is the gaussian fit to the background.}
         \label{bgfit}
   \end{figure}

\begin{figure}
   \centering
   \includegraphics[width=8 cm]{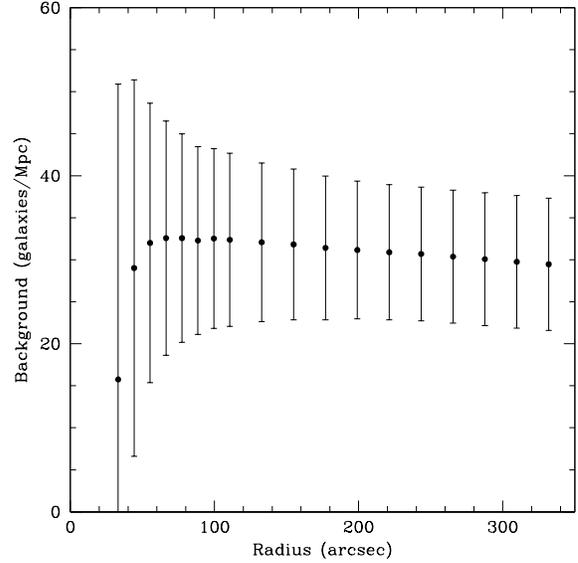}
      \caption{Background mean and error versus smoothing radius for the full catalogue of cluster members. For very low values (under 60") the value rapidly goes down due to oversampling of small fluctuations, while beyond 120" a trend towards lower values is due to undersampling of counts at the field border. Between 60" and 120", the background value is stable. Error bars are 1$\sigma$ errors.}
         \label{bgerrors}
   \end{figure}

To select the most efficient smoothing radius, several test were run on the sample with different radius values. The background was thus calculated with increasing radii and then fitted with a gaussian, assuming it is dominated by random noise (i.e. randomly scattered field galaxies). The background mean value and RMS are weakly sensitive to the smoothing radius for radius values between 60" and 120". Below 60", small-scale fluctuations make the background dominated by small numbers, while beyond 120" the mean goes slowly but steadily down, due to undersampling of background counts at the field borders. Between 60" and 120" the background value is almost stable, with errors becoming slightly lower with increasing smoothing radius. The behaviour, along with the background RMS, is shown in Fig. \ref{bgerrors}. A final smoothing scale of 110" was chosen; this is equivalent to 500 kpc at the cluster redshift.

After evaluating background mean and RMS, the true density maps are generated.
First, galaxies are counted in each mesh box and their number is divided by the box area and then normalised to Mpc scale. The map is then smoothed with a gaussian filter to clean numerical spikes (arising at the map border due to wrong sampling) and for better, continuous connection of counts in neighboring mesh cells. The same smoothing radius as for background calculation gives reliable results, cleaning spikes without deleting substructuring information.
A number of simulated background maps is then generated as randomly distributed values with normal distribution (as given by the background best-fit parameters and chosen smoothing scale); the background map is then independently subtracted from the raw density map and the resulting background-subtracted maps are stacked together and normalised. This ensures a background subtraction as smooth as possible and unaffected by true overdensities in the cluster field. 100 iterations already give a good mean background subtraction.

The background-subtracted, stacked and normalised image is in the end taken as the final surface density map of the field.

\end{appendix}

\begin{appendix}
\section{Spectroscopic catalogue of RXCJ0014.3-3022 (A2744)}
Hereafter the full catalogue of spectroscopic objects in the field of RXCJ0014.3-3022 is provided. Along with position and redshift, for each object are also provided the values of spectral indices and BVR magnitudes with corresponding errors. Where equivalent widths were not calculated due to the line not being sampled in the instrument's spectral range, the table reads ND, i.e. Not Detected.

\end{appendix}

\begin{appendix}
\label{photcat}
\section{Spectroscopic catalogue of RXCJ2308.3-0211 (A2537)}
Hereafter the full catalogue of spectroscopic objects in the field of RXCJ2308.3-0211 is provided. Along with position and redshift, for each object are also provided the values of spectral indices and BVR magnitudes with corresponding errors. Where equivalent widths were not calculated due to the line not being sampled in the instrument's spectral range, the table reads ND, i.e. Not Detected.

\end{appendix}

\end{document}